\documentclass[
 reprint,
 amsmath,amssymb,
 prb,
 twocolumn,
superscriptaddress
]{revtex4-2}
\usepackage{mystyle}
\usepackage{graphicx}
\usepackage{dcolumn}
\usepackage{bm}
\usepackage{upgreek}
\usepackage{subfigure} 

\usepackage{xcolor}

\definecolor{darkyellow}{RGB}{246,201,68}

\def\rs{r_{{\rm{s}}}}
\def\Smc{\mathcal{S}}
\def\Gbf{\mathbf{G}}
\def\gel{g}
\def\U{\mathrm{U}}
\def\Cmc{\mathcal{C}}
\def\Amc{\mathcal{A}}
\def\Xmc{\mathcal{X}}
\def\nel{n_{\rm{el}}}
\def\ua{\uparrow}
\def\vbs{\boldsymbol{v}}
\def\rbs{\boldsymbol{r}}
\def\da{\downarrow}

\def\Xsf{\mathsf{X}}
\def\Ysf{\mathsf{Y}}

\def\Hsf{\mathsf{H}}
\def\Ssf{\mathsf{S}}

\def\zero{\mathbf{0}}
\def\SP{\rm{SP}}
\def\VP{\rm{VP}}

\def\qbs{\mathbf{q}}
\def\kbs{\mathbf{k}}
\def\ep{\varepsilon}

\def\ii{{\rm{i}}}
\def\Area{{\rm{A}}}
\def\Hsf{\mathsf{H}}
\def\Asf{\mathsf{A}}

\def\pbs{\mathbf{p}}
\def\Isf{\mathsf{I}}
\def\Usf{\mathsf{U}}
\def\Gmc{\mathcal{G}}

\def\Xmc{\mathcal{X}}

\def\Smc{\mathcal{S}}
\def\pSP{{\rm{pSP}}}
\def\pVP{{\rm{pVP}}}

\begin{document}


\title{
Symmetry-determined generalized ferromagnetism in multi-valley electron fluids
}

\author{Vladimir Calvera}
\affiliation{Department of Physics, Stanford University, Stanford, CA 93405}
\affiliation{Kavli Institute for Theoretical Physics, University of California, Santa Barbara, California 93106, USA}

\author{Erez Berg}
\affiliation{Department of Condensed Matter Physics, Weizmann Institute of Science, Rehovot 7610001, Israel}

\author{Steven A. Kivelson}
\affiliation{Department of Physics, Stanford University, Stanford, CA 93405}

\date{\today}

\begin{abstract}
Quantum electronic fluids with spin and valley degrees of freedom have a correlation driven tendency to flavor polarization (generalized ferromagnetism). To first order in the long-range Coulomb interactions -- i.e. in the Hartree-Fock approximation -- spin and valley polarization exhibit a spurious degeneracy.
We show that to second order -- or more generally in the random-phase approximation -- this degeneracy is lifted in a way that depends only on the underlying symmetry relating the two valleys. In two spatial dimensions, if the valleys are related by an $n-$fold rotation ($n>2$) or by mirror reflection and each valley is invariant under $C_2$ or time reversal  
(as is the case in AlAs quantum wells) then valley polarization is preferred. 
If the valleys are related by time reversal or by $C_2$ rotation symmetry (as in multilayer graphene systems) then spin order is selected. 
\end{abstract}

\maketitle

\textit{Introduction.--} 
Strongly correlated electron fluids host a rich variety of emergent phenomena.  As an appropriate dimensionless measure
\footnote{For different forms of the interaction - e.g. various screening gate geometries, and different electron dispersions one may have to adopt somewhat different methods to define $r_{\rm{s}}$.  Typically, $r_{\rm{s}}$ increases with decreasing electron density. For Coulomb interactions and two-dimensional electrons with quadratic dispersion relation, $r_{\rm{s}}\equiv \frac{m e^2}{4\pi \epsilon \hbar^2\sqrt{\pi \nel}}$, where $\nel$ is the electron density, $m$ is the band mass of the electron, and $\epsilon$ is the dielectric constant of the medium.
}
of the strength of the electron-electron interactions, $r_{\rm{s}}$, become sufficiently large, the system often becomes unstable to forming some type of electronic order.
One of the simplest such instabilities is toward itinerant ferromagnetism~\cite{stoner1938collective}. In the canonical homogeneous two-dimensional electron gas (2DEG), however, this tendency appears to be preempted by Wigner crystallization~\cite{Tanatar19892DEG_QMC, Attaccalite2002DEG_qMC,Drummond20092DEG_DMC,Azadi2024QMC_2DEG,Smith20242DEG_NN_QMC,falson2022competing}.
\begin{figure}[h!]
    \centering
    \includegraphics[width=0.85\linewidth]{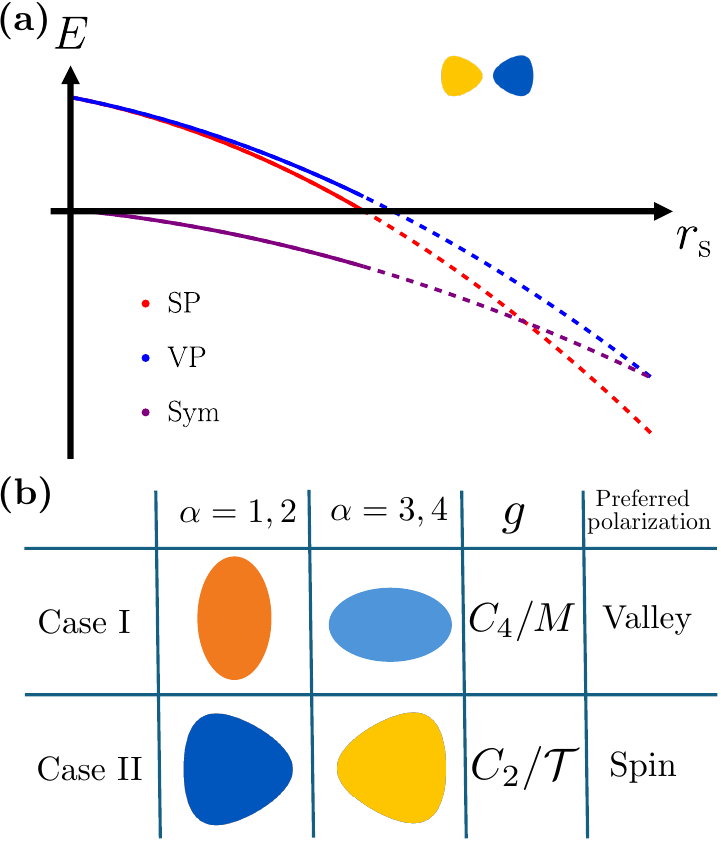}
    \caption{
    \textbf{(a)} Schematic evolution of the energy density of the valley polarized (VP), spin polarized (SP) and symmetric (Sym) states as a function of the interaction strength ($r_{\rm{s}}$)
    {for a pair of valleys related by $\qbs \to -\qbs$}  (Case II, below).
    Solid lines indicate controlled calculations; dashed lines are extrapolations. 
    \textbf{(b)}  Preferred form of flavor polarization for different symmetry Fermi pockets. Shapes indicate typical Fermi seas for electrons with flavor $\alpha$. $\gel$ are examples of symmetries relating different valley/flavors. $C_n$ are in-plane $n$-fold rotations, $\mathcal{T}$ is time-reversal, and $M$ is a reflection through a plane perpendicular to the two-dimensional electron system. A multi-valley system where the valleys are related by $C_n$  with $n>2$ behaves similarly to case I.
    }
    \label{fig:1}
\end{figure}
More recently, attention has turned to multi-valley electron gases—systems in which the conduction band 
has multiple degenerate minima (valleys) in momentum space, endowing electrons with an additional flavor degree of freedom. In such systems, generalized 
ferromagnetism, involving spontaneous spin and/or valley polarization, 
can emerge prior to crystallization~\cite{Nematic2DEG_VMC,Hossain2021AlAsValleyPolarization,zhou2021half,zhou2021superconductivity,de2022cascade,liu2024spontaneous,angyan2011correlation}.

The dominant interaction in these systems is typically the  long-ranged density-density (Coulomb) repulsion, which is SU$(2n_v)$ symmetric in spin and valley space, where $n_v$ is the number of valleys. This symmetry implies a degeneracy among different types of generalized ferromagnetic orders: spin polarization, valley polarization, or mixed configurations. However, 
 the valleys are 
not identical.
The electronic dispersion relation of each valley is generally not invariant under the action of the subset of crystalline symmetries that relates the valleys. 
Taking this into account reduces  the SU$(2n_v)$ symmetry down to 
the actual crystalline symmetries, and as a consequence 
selects a preferred type of flavor order.

Theoretical studies of flavor polarization in such systems are commonly performed within the Hartree-Fock (HF) approximation, which captures exchange-driven instabilities. While HF predicts spontaneous flavor polarization at sufficiently large $r_s$, it also tends to overestimate ordering tendencies and often exhibits accidental degeneracies between states not related by any symmetry. For instance, in the presence of purely long-range Coulomb interactions, HF yields identical variational energies for spin-polarized and valley-polarized states. 

In this work, we investigate how these accidental degeneracies are lifted by correlation effects. 
Using second order perturbation theory and the random phase approximation (RPA), we elucidate 
the physics governing the most favored form of flavor polarization. 
As illustrated in Fig.~\ref{fig:1}a, 
the approach is controlled in the weak coupling (small $\rs$) limit, but when extrapolated to intermediate couplings, it leads to a plausible determination of the nature of the flavor ferromagnetism expected beyond a critical value, $\rs > \rs^c$.  {(It has been shown previously}~\cite{ZhangDasSarma2005RPA_transitions,jang2023chirality,calvera2024theory} that the RPA significantly improves - but does not  eliminate - the tendency of HF to 
underestimate 
$\rs^c$.)

Our principle conclusion is that the 
energetically favored form of generalized ferromagnetism depends only on the symmetry that relates the valleys.  This is illustrated in Fig.~\ref{fig:1}b for the case of two valleys and no spin-orbit coupling.  The upper row illustrates a class of band structures, of which AlAs heterostructures are an example, in which the dispersion relation in each valley is time-reversal symmetric, while a $C_4$ rotation interchanges the two valleys - in this case, the valley polarized, spin unpolarized state is favored.  The lower row shows a case in which the two valleys are interchanged under time-reversal, as in multilayer graphene systems - in this case, the spin polarized, valley symmetric state is  favored.

{ Our analysis treats only the dominant long-ranged part of the Coulomb interaction. We neglect
the large momentum transfer (intervalley) matrix elements of the interaction -- which are suppressed by a factor of order $\sqrt{\nel}a$, where $\nel$ is the electron density and $a$ is the microscopic lattice spacing of the host material.}
We also neglect spin-orbit coupling, i.e. we  assume spin rotation invariance. %
In some cases, even though these terms are relatively small, they can lift remaining degeneracies and hence play a role in determining the precise form of generalized ferromagnetism, as discussed in App.~\ref{app:PerturbationsSP}.

\textit{Setup.--} 
We consider the following second-quantized Hamiltonian of fermions in two dimensions
\begin{equation}\label{eq:Ham}
\begin{split}
    H &= \sum_{\kbs,\alpha}\ep_{\kbs,\alpha} c_{\kbs,\alpha}^{\dagger}c_{\kbs,\alpha}^{\,}+ \frac{1}{2\Area} \sum_{\qbs \neq \mathbf{0}} v_{\qbs} \frac{:\rho_{\qbs}^{\dagger}\rho_{\qbs}+\rho_{\qbs}\rho_{\qbs}^{\dagger}:}{2},  \\
    \rho_{\qbs} &= \sum_{\kbs,\alpha} \Lambda_{\kbs+\qbs,\kbs}^{\alpha}c_{\kbs+\qbs,\alpha}^{\dagger}c_{\kbs,\alpha},
\end{split}
\end{equation}
where $v_{\qbs}=v_{-\qbs}\geq 0$ is the Fourier transform of the interaction, $c_{\kbs,\alpha}^{\dagger}$ are creation operators, $:\cdots:$ denotes normal ordering, $\Area$ is the area, $\varepsilon_{\kbs,\alpha}$ is the electronic dispersion, and $\Lambda_{\kbs',\kbs}^{\alpha} = \braket{u_{\kbs',\alpha}}{u_{\kbs,\alpha}}$ is the `form factor,' defined as the overlap of the periodic parts of the Bloch eigenstates, denoted as $\ket{ u_{\kbs,\alpha}}$. 
For simplicity, we treat the case of two valleys ($n_v=2$), although the results easily generalize to systems with a larger number of valleys.
The flavor index, $\alpha$, takes on four values, with $\alpha=1,2,3,4$ corresponding to the states 
{ $\{\tau=1,s=\ua\}, \{\tau=1,s=\da\},\{\tau=3,s=\ua\},\{\tau=4,s=\da\}$}
, respectively, where $\tau$ and $s$ are the valley and spin indices. 

We choose a gauge such that the eigenstates satisfy $\ket{u_{\kbs,1}} = \ket{u_{\kbs,2}}$ and $\ket{u_{\kbs,3}} = \ket{u_{\kbs,4}}$. Beyond the constraints of symmetry, our results hold independent of the specific form of $\varepsilon_{\kbs,\alpha}$, $\Lambda^\alpha_{\kbs,\kbs'}$, and $v_{\qbs}$, but to have a specific example in mind, one can imagine it to be the Coulomb interaction screened by equidistance metallic gates $v_{\qbs} = \frac{e^2}{2q }\tanh(qd)$, where $d$ is the sample-to-gate distance.

{The Hamiltonian $H$ has a $[U(2)\times U(2)]\rtimes g$ } 
 symmetry, where each $U(2)$ corresponds to spin and charge conservation on a valley, and  there is an  additional discrete symmetry that exchanges the two valleys, that we will refer to as
$g$. { Note that $g$ could be a point group symmetry or time-reversal symmetry. 
We refer to any symmetry such that $g[\qbs] = -\qbs$ as a `momentum-negating symmetry'. Examples of momentum negation symmetries are time-reversal symmetry ($\mathcal{T}$) and spatial inversion ($\mathcal{I}$) { (which is equivalent to 
$C_2$, 
a two-fold rotation, for two-dimensional systems)}. However, their combination ($\mathcal{TI}$) is not momentum-negating. }

Defining $\qbs'\equiv g[\qbs]$, the discrete symmetry implies: $v_{\qbs'}=v_{\qbs}$, $\ep_{\kbs,\alpha+2}=\ep_{\kbs',\alpha}$
, and $\Lambda_{\kbs+\qbs,\kbs}^{\alpha+2}= \Lambda_{\kbs'+\qbs',\kbs'}^{\alpha}$ or $\Lambda_{\kbs+\qbs,\kbs}^{\alpha+2}= \Lambda_{\kbs',\kbs'+\qbs'}^{\alpha}$ depending on whether $g$ is unitary or anti-unitary. (Addition of flavor labels is defined modulo 4.) To simplify the discussion, we assume that $g^2$ is a symmetry of each flavor.
We consider two cases, illustrated in Fig.~\ref{fig:1}b:
\begin{enumerate}
    \item \textit{Case I:} The two valleys are invariant under momentum-negation and related to each other by another symmetry. For example, this is the case for two valleys with anisotropic parabolic dispersions $\ep_{\kbs,\alpha} =\frac{\kbs^{T} M_{\alpha}^{-1}\kbs}{2}$, arising from band minima at the Brillouin zone (BZ) edges, as e.g. in AlAs quantum wells \cite{Hossain2021AlAsValleyPolarization, shayegan2006two}.
    \item \textit{Case II:} The two valleys are related by momentum-negation, and each valley is \textbf{not} invariant under momentum-negation. For example, this is the case for valleys from band minima at the corners of the hexagonal BZ ($K$ and $K'$ points) when the effect of trigonal warping is included. This is relevant for experiments in multi-layer graphene systems in the presence of a displacement field \cite{zhou2021half,de2022cascade,liu2024spontaneous}. 
\end{enumerate}

For weak interactions, the ground state has an equal number of electrons of each flavor to minimize the kinetic energy. As interactions get stronger, electrons can gain exchange energy by spontaneously developing flavor polarization. However, another way to gain energy is by creating correlations beyond {those imposed by} Fermi statistics. 
Predicting whether the ground state is flavor polarized for a specific electron density is a notoriously hard problem. Rather than attempting to solve it, we 
ask: \textit{What is the most likely type of flavor polarization?}

As a first step, one may obtain the (restricted) Hartree-Fock phase diagram by comparing the energy of different Slater determinants that represent the competing phases. 
One important set of such phases are ``half-metals'', in which electrons occupy only half of the flavors. 
By symmetry, it suffices to consider two half-metals: 1) the spin-polarized liquid (SP); 2) the valley-polarized liquid (VP). In SP (VP) only orbitals with $\alpha=1,3$ ($\alpha=1,2$) are occupied, respectively. 
Because of the $\U(2)\times \U(2)$ symmetry of the Hamiltonian, SP refers to the family of valley-unpolarized states that are spin ferromagnets in each valley with arbitrary spin-orientation between valleys; the ferromagnet is actually a point on a $S^2\times S^2$ manifold of exactly degenerate states.

{
Within HF, the SP and VP states are product states: $\ket{\Psi_{\VP}} = C_{1}^\dagger C_{2}^\dagger \ket{\Omega} $ and $\ket{\Psi_{\SP}} = C_{1}^\dagger C_{3}^\dagger \ket{\Omega} $, where $\ket{\Omega}$ is the vaccum, $C_\alpha^{\dagger}:= \prod_{\kbs: \varepsilon_{\kbs}^{\alpha} < E_{F}} c_{\kbs,\alpha}^{\dagger}$ and $C_{3}^{\dagger} = U\cdot C_2^{\dagger}\cdot  U^{\dagger}$, where $U$ is the operator implementing the symmetry $c^{\dagger}_{\kbs,(\sigma,\tau)}\to c^{\dagger}_{g[\kbs],(-\sigma,g[\tau])}$ \footnote{Here $E_F$ is the Fermi energy chosen to have the correct electron density. }. 
It is straightforward to show that $\mel{\Omega}{C_{\alpha}^{}C_{\alpha'}^{}H C_{\alpha}^{\dagger}C_{\alpha'}^{\dagger}}{\Omega} =  \mel{\Omega}{C_{\alpha}^{}H C_{\alpha}^{\dagger}}{\Omega} + \mel{\Omega}{C_{\alpha'}^{}H C_{\alpha'}^{\dagger}}{\Omega}$, so the HF energies of the half-metals can be written as $E^{\rm{HF}}_{\VP} = E^{\rm{HF}}_1+ E^{\rm{HF}}_2$ and $E^{\rm{HF}}_{\SP} = E^{\rm{HF}}_1+ E^{\rm{HF}}_3$, where $E^{\rm{HF}}_\alpha = \mel{\Omega}{C_{\alpha} H C_{\alpha}^\dagger}{\Omega}$. As $C_3^\dagger=UC_2^\dagger U^\dagger$ and $UH=HU$, it follows that $E_2^{\rm{HF}}=E_3^{\rm{HF}}$, i.e. the Hartree-Fock energies of SP and VP are equal. 

}

Before proceeding, we introduce the (non-interacting) polarization bubbles, as they play a central role in our analysis. The polarization bubble, $\Pi(\qbs, \ii\omega)$, are the non-interacting susceptibility between $\rho_{\qbs}$ and $\rho_{\qbs}^\dagger$.
For product {states 
as above the susceptibility reduces to $\Pi = \sum_{\alpha}\Pi_{\alpha}$ where $\Pi_{\alpha}$ is the susceptibility of the state 
${C}_{\alpha}^\dagger\ket{\Omega}$
, and the product runs over the occupied flavors.} These are given by
\begin{equation}\label{eq:bare_bubble}
    \Pi_{\alpha}(\qbs,\ii\omega) = \frac{1}{\Area}\sum_{\kbs} \frac{n_{\kbs+\qbs,\alpha}-n_{\kbs,\alpha}}{\ii\omega-(\ep_{\kbs+\qbs,\alpha}-\ep_{\kbs,\alpha})}\abs{\Lambda_{\kbs+\qbs,\kbs}^{\alpha}}^2,
\end{equation}
where $n_{\kbs,\alpha}$ is the occupation number. When $\Pi_{\alpha}$ appears with different values of $\alpha$ in the same set of equations, all are evaluated using a common Fermi energy, $E_F^{\alpha} = E_F$.

The polarization bubbles have the following properties: 
(1) $[\Pi_{\alpha}(\qbs,\ii\omega)]^*=\Pi_{\alpha}(-\qbs,\ii\omega)=\Pi_{\alpha}(\qbs,-\ii\omega)$; (2) If the state is invariant under $\gel$, then $\Pi_{\gel(\alpha)}(\qbs,\ii\omega)=\Pi_{\alpha}(\gel(\qbs),\ii\omega)$; (3) The real part of $\Pi_{\alpha}$ is non-negative. (See App.~\ref{app:BubbleProps} for a derivation.)

\textit{Second order perturbation theory -}{
To determine the lifting of the degeneracy between the SP and VP states, we examine second-order corrections to their energies. Most terms in second-order perturbation theory (2PT) are independent of the pattern of flavor polarization. 
The contribution that distinguish SP and VP is
\begin{equation}\label{eq:2PT:ringPrimed}
    E_{\rm{r}}' = - \frac{1}{2}\Re\left[\int_{\qbs,\omega}v_{\qbs}^2 \ [\Pi_{\alpha'}(-\qbs,\ii\omega)]^*\Pi_{1}(\qbs,\ii\omega)
    \right],
\end{equation}
where $\int_{\qbs,\omega} \equiv \frac{1}{\Area} \sum_{\qbs}\int_{\mathbb{R}}\frac{\dd\omega}{2\pi}$, and $\alpha'$ is the second occupied flavor, i.e. $\alpha'=2$ for VP, and $\alpha'=3$  for SP. (See App.~\ref{app:2PT} for more details.)

For complex functions of momentum  and frequencies $f$ and $g$, $(f,g) \equiv \int_{\qbs,\omega}v_{\qbs}^2f(\qbs,\omega)^*g(\qbs,\omega) $ defines an inner product. Then, $E'_{\rm{r}} = -\frac{1}{2}(\Isf[\Pi_{\alpha'}],\Pi_{1})$, where $\Isf[f](\qbs,\omega)=f(-\qbs,\omega)$. Due to symmetry $\Pi_{\alpha'} = \Usf_{\alpha'}[\Pi_{1}] $ for some unitary $\Usf_{\alpha'}$ acting solely on the momentum index. Thus the Cauchy-Schwartz inequality implies that $E_{\rm{r}}' \geq -\frac{1}{2}(\Pi_1,\Pi_1)$, with equality only attained when $\Pi_{\alpha'}(\qbs,\ii\omega) = \Pi_1(-\qbs,\ii\omega)$. This implies the results in Fig.~\ref{fig:1}b because in Case I, all bubbles are invariant under momentum inversion by assumption ($\Pi_{\alpha}=\Isf[\Pi_{\alpha}]$) and $\Pi_{1} = \Pi_{2} \neq \Pi_{3}$. In contrast for Case II, $\Pi_{1} = \Isf[\Pi_{3}] \neq \Pi_2$.

To gain some intuition, we can express $E'_{\rm{r}}$ as 
\begin{equation}
    E'_{\rm{r}} = -\sum_{\qbs}  \frac{v_{\qbs}^2}{2\Area}
    \sum_{\substack{i\in \rm{PH}(+\qbs,1) \\ j\in \rm{PH}(-\qbs,\alpha')}}
    \frac{\abs{\Lambda_i}^2\abs{\Lambda_j}^2}{\varepsilon_i + \varepsilon_j},
\end{equation}
where $\rm{PH}(\qbs,\alpha)$ is the set of particle-hole excitations with momentum $\qbs$ created by operators $b_{j}^\dagger \sim c_{\kbs+\qbs,\alpha}^{\dagger}c_{\kbs,\alpha}^{}$. The particle-hole pair $j$ has energy $\varepsilon_i = \varepsilon_{\kbs+\qbs,\alpha} - \varepsilon_{\kbs,\alpha}>0$ and matrix element $\Lambda_j = \Lambda_{\kbs,\kbs+\qbs}^{\alpha}$.

The difference in energy between different flavor polarizations thus comes from differences in the particle-hole spectra of the occupied flavors, and it is most favorable to have the particle-hole spectrum of the occupied flavors to be related by momentum-negation. 
}

\textit{Random Phase Approximation.--}  
For long-ranged Coulomb interactions, perturbation theory contains infrared divergences. These divergences are cured by resumming perturbation theory within the RPA~\cite{GellMann,Rajagopal1977RPAResummation2DEG}. The RPA correlation energy can be calculated by coupling-constant integration \footnote{This approximation is equivalent to using the Hellmann-Feynman theorem combined with the time-dependent Hartree approximation (TDH) for correlation functions
(See App.~\ref{app:CouplingConstant:FH}).
}, and is given by
\begin{equation}\label{eq:CorrelationEnergy}
    E_{\rm{c}} = 
    \sum_{\qbs}\frac{v_{\qbs}}{2\Area} 
    \int_{0}^{1}\left(\bar{\Smc}(\qbs;\lambda)-\bar{\Smc}(\qbs;0)\right)\dd{\lambda},
\end{equation}
where $\bar{\Smc}(\qbs;\lambda)$ is given by
\begin{equation}\label{eq:StructureFactor}
    \bar{\Smc}(\qbs;\lambda) = \int_{-\infty}^{\infty} \Re\left[\frac{\chi_0(\qbs,\ii\omega)}{1+\lambda v_{\qbs} \chi_0(\qbs,\ii\omega)} \right]\frac{\dd{\omega}}{2\pi}.
\end{equation}
$\chi_0$ is the non-interacting charge susceptibility. For VP and SP states, $\chi_0^{\VP} = \Pi_1+\Pi_2=2\Pi_1$ and $\chi_0^{\SP} = \Pi_1 + \Pi_3$, respectively. 

As $\bar{\Smc}(\qbs;0)$ is identical for SP and VP, we only need to compare $\bar{\Smc}(\qbs;\lambda)$ for both polarizations to identify which one is favored by correlation effects. To this end, we focus on the difference $\Delta \bar{\Smc}(\qbs) \equiv  \bar{\Smc}_{\VP}(\qbs)-\bar{\Smc}_{\SP}(\qbs)$.

For case I, the invariance of each valley under $\qbs \to -\qbs$, implies that the polarization bubbles are real and non-negative. Consider the non-negative functions $\Xsf_1 \equiv v \Pi_{1}$ and $\Xsf_2 \equiv v \Pi_{3}$.
The product of the potential energy and the bare charge susceptibility, $v\chi_0$, for SP and VP are $[v\chi_{0}]_{\SP}= \Xsf_1 +\Xsf_2$ and $[v\chi_{0}]_{\VP}= 2\Xsf_1$. By the strict concavity of the function $x\mapsto x/(1+x)$, we have 
\begin{equation}
    \frac{2\Xsf_1}{1+2\Xsf_1} + \frac{2\Xsf_2}{1+2\Xsf_2}
\leq 2\frac{(\Xsf_1+\Xsf_2)}{1+(\Xsf_1+\Xsf_2)},
\end{equation}
where the equality holds only when $\Xsf_1=\Xsf_2$. The latter inequality implies
\begin{equation}\label{eq:CaseI:Ineq}
    \bar{\mathcal{S}}_{\VP}(\qbs;\lambda) + \bar{\mathcal{S}}_{\VP}(\gel^{-1}(\qbs);\lambda)\leq 2 \bar{\mathcal{S}}_{\SP}(\qbs;\lambda) .
\end{equation}
Summing over $\qbs$ and $\gel(\qbs)$, we obtain $\Delta \mathcal{S}(\qbs;\lambda)+ \Delta \mathcal{S}(\gel(\qbs);\lambda) \leq 0$, which implies that $\Delta E \leq 0$. In other words, VP has lower correlation energy than SP, independent of most details of the electronic structure.
We reached this conclusion in our previous work \cite{calvera2024theory}.

For case II, the polarization bubbles are complex, but we can express $\Delta \Smc$ in terms of two real functions: $\Xsf \equiv v (\Pi_{1}+\Pi_{3})\geq 0$ and $\Ysf \equiv v (\Pi_{1}-\Pi_{3})/\ii$. 
Then, $[v\chi_{0}]_{\SP}= \Xsf$ and $[v\chi_{0}]_{\VP}= \Xsf + \ii \Ysf$. After some algebra, 
\begin{equation}\label{eq:CaseII:Ineq}
    \Delta \bar{\mathcal{S}} = \int_0^{\infty} \frac{\Ysf^2}{(1+\Xsf)((1+\Xsf)^2+\Ysf^2)}\frac{\dd\omega}{\pi} \geq 0
\end{equation}
Thus SP has lower energy than VP in case II.

{Note that because the imaginary part of $\chi_0(\qbs,\ii\omega)$ is odd in frequency, $\chi_0(\qbs,0)$ is identical for SP and VP. Therefore, the difference in the correlation energy between the VP and SP half-metals is missed in the commonly used approximation where correlation effects are included by replacing the bare interaction with the static RPA screened interaction $v_{\qbs}/(1+v_{\qbs}\chi_0(\qbs,0))$.

{
Often, 
a significant part of the correlation energy can be associated with the zero-point energy plasmon excitation as the density fluctuation spectrum is dominated by the plasmon pole at long wave-lengths~\cite{Giuliani_Vignale_2005_chapter5}.
It is therefore interesting to examine the plasmon frequencies of the different states. 
We calculated the 
plasmon dispersions for small $\abs{\qbs}$ \footnote{See App.~\ref{app:Plasmon} for details.}. In both cases, we found that $\bar{E}_{\rm{VP}}(\qbs)\leq  \bar{E}_{\rm{SP}}(\qbs)$, where $\bar{E}(\qbs) = E_{\rm{pl}}(\qbs) + E_{\rm{pl}}(g[\qbs])$ and $E_{\rm{pl}}(\qbs)$ is the energy of the plasmon with momentum $\qbs$. 
This shows that plasmons do not dominate the SP–VP energy difference in Case II.}

The results obtained above can be generalized in several interesting ways. 
In App.~\ref{app:Extensions}, we show that within RPA, the trends summarized in Fig.~\ref{fig:1}b persist in {at least} the following cases:
(1) A multi-component system, e.g., a multi-layer system, where the interaction is layer dependent \footnote{In a multilayer system, each $\rho_{\qbs, \mu}$ is $\qbs$ Fourier mode of the density projected to layer $\mu$ and $[v_{\qbs}]_{\mu\nu} \sim \frac{e^2}{2\epsilon \abs{\qbs}}e^{- d_{\mu\nu}\abs{\qbs}}$ with $d_{\mu\nu}$ the distance between layers $\mu$ and $\nu$.}.
(2) A three-dimensional system. 
(3) A system with an orbital magnetic field. 
In addition, we show in App.~\ref{app:Extensions} that the results persist if we use  time-dependent Hartree-Fock approximation (TDHFA), instead of TDH in calculating $\bar{\Smc}(\qbs)$.
}

\begin{figure}[h]
    \centering
    \includegraphics[width=0.95\linewidth]{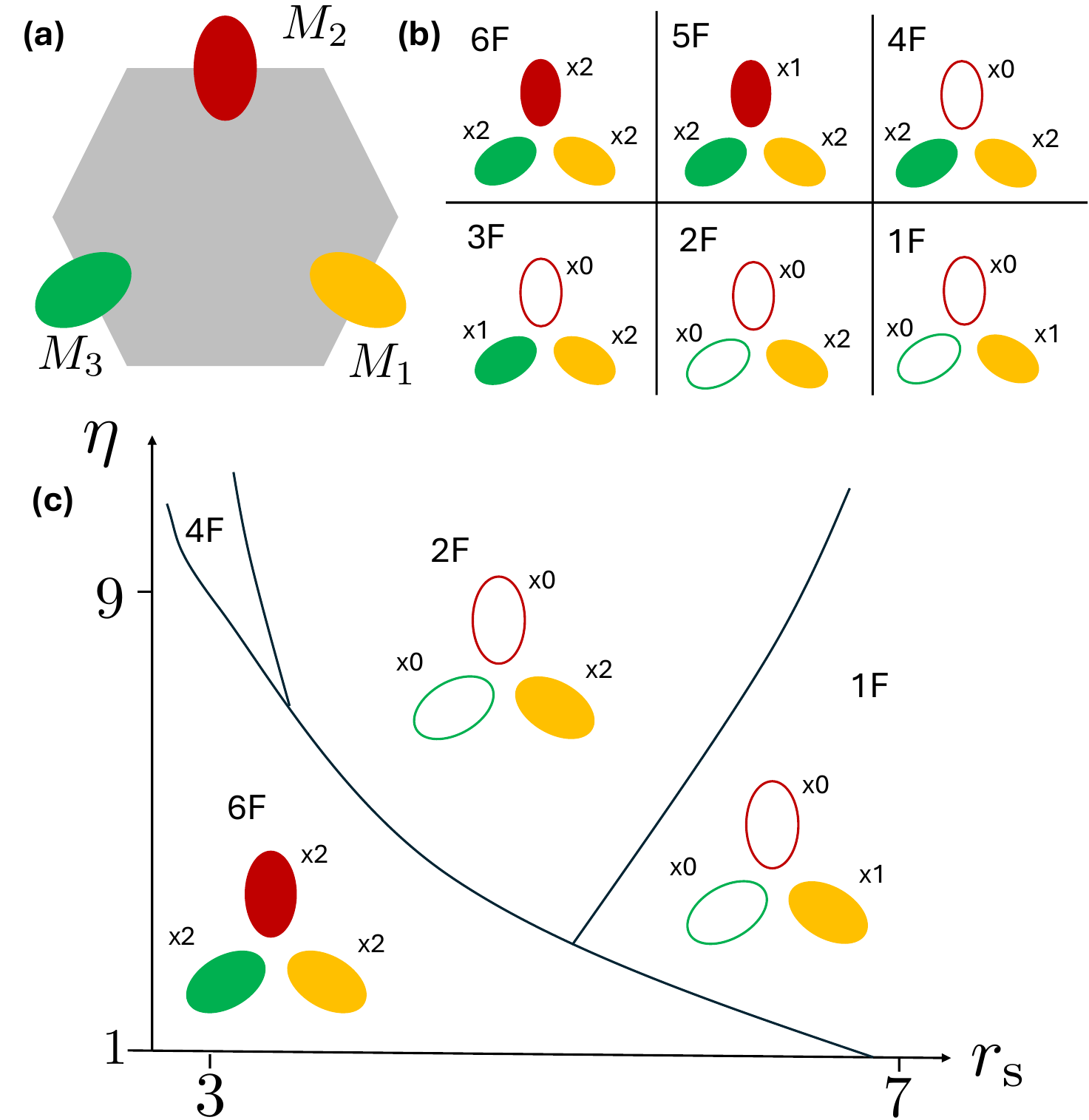}
    \caption{ \textbf{(a)} Schematic band structure of three valleys centered at the $M$ points of the Brillouin zone, related by $C_3$ rotations.
    \textbf{(b)} Most likely flavor polarization patterns for metallic states with equally populated flavors as predicted by RPA. Each label `nF' denotes a metal with `n' equally occupied flavors. 
    The spin multiplicity associated with each valley is indicated.
    \textbf{(c)} Conjectured phase diagram for a parabolic dispersion as a function mass anisotropy $\eta$ { (defined as the ratio of the masses along the principal axes) } and interaction strength $r_s$, motivated by the RPA results and the explicit calculation for the model with two valleys \cite{Nematic2DEG_VMC,calvera2024theory}.
    }    
    \label{fig:fig_M_pts}
\end{figure}

\textit{Partial half-metals and susceptibilities.--} We can also consider polarized state such that the density of electrons with each flavor is $\nel(1\pm\zeta)/4$, where $\zeta \in (0,1)$ characterizes the degree of polarization and $\nel$ denotes the total electron density. A partial VP (SP) state corresponds to the case where flavors 1 and 2 (3) have density $\nel(1+\zeta)/4$. The same arguments used for the the half-metals show that the partial VP (SP) is favored in Case I (II). 
This result allows us to conclude that the susceptibilities of the symmetric (unpolarized) state follow the same trends—for example, the valley susceptibility is larger than the spin susceptibility in Case I.

\textit{Three valleys related by $C_3$ ($n_v=3$).--} Our results naturally extend to a $C_3$ symmetric system with Fermi pockets centered at the $M$ points of the hexagonal Brillouin zone (Fig.~\ref{fig:fig_M_pts}a).  Each of the three distinct $M$ point vallies is invariant under momentum-negation. This situation could be relevant for the monolayer transition metal dichalcogenides 1T-HfX$_2$ (X=Se,S) \cite{zhao2017elastic,yan20182d,keshri2023hfse2,PhysRevB.98.155204}. 

There are now six flavors of fermions (three valleys and two spins). 
HF predicts that all states with $n=2,3,4,5$ of the equally occupied and $6-n$ unoccupied flavors are degenerate. 
This degeneracy is lifted by correlation effects. A calculation of the correlation energy within 2PT or RPA following the lines of Eq.~(\ref{eq:2PT:ringPrimed}-\ref{eq:CaseI:Ineq}) predicts that states that maximize the valley polarization are favored (see Fig.~\ref{fig:fig_M_pts}b) \footnote{{The assumption of equal size Fermi surfaces for all occupied flavors is natural for all states other than 3F and 5F;  however, these two states are
    likely unstable to the transfer of electrons from the spin-polarized valley to the spin-degenerate ones.}}. For example, in the 2F state, a single valley is occupied with both spin flavors; in the 4F state, two valleys are occupied an one is completely empty.  
These considerations, as well as insights gained from detailed RPA~\cite{calvera2024theory} and VMC~\cite{Nematic2DEG_VMC} calculations for the case with $n_v=2$, allow us to construct a schematic conjectured phase diagram of the system as a function of the effective mass anisotropy $\eta$ of each valley and the interaction strength, parametrized by $r_s$ (Fig. \ref{fig:fig_M_pts}c). The detailed arguments leading to this phase diagram are provided in App.~\ref{app:EndMatter:ThreeValleys}. 

\textit{Conclusion.--} 
We have shown that in multi-valley electron fluids, correlation effects beyond Hartree-Fock lift many of the accidental degeneracies between distinct generalized ferromagnetic states. 
Our central finding is that the preferred pattern of spin and valley polarization is dictated by the symmetry relating the valleys. Specifically, configurations in which the dispersions of occupied flavors are connected by momentum-negating symmetries gain the most correlation energy.

This conclusion is most transparent within second-order perturbation theory [Eq.~\eqref{eq:2PT:ringPrimed}], where the correlation energy involves an integral over a product of polarization bubbles. When two occupied flavors are related by momentum negation, their polarization functions optimally overlap within this integral, maximizing the correlation energy gain. The same conclusions apply within RPA. 
More accurate VMC calculations for the case where the valleys are related by $C_4$ further support these results~\cite{Nematic2DEG_VMC}; VMC calculations for the other case, where the valleys are related by momentum negation, as well as for the $C_3$ symmetric case, are highly desirable. 

Physically, the difference in energy
gain arises from virtual particle-hole excitations with opposite momenta in distinct flavors -- i.e., from flavor density wave fluctuations. These fluctuations act analogously to the van der Waals interaction between non-polar molecules: just as induced dipoles in polarizable directions reduce the interaction energy, density waves in different flavors interact most favorably when the softest modes are matched in momentum and frequency. 

Our results naturally explain why in AlAs quantum wells, where the two valleys are related by $C_4$, a valley-polarized, spin-unpolarized state is observed~\cite{Hossain2021AlAsValleyPolarization}. In contrast, to the best of our knowledge, in rhombohedral graphene multilayers, all the observed half-metal states are either spin-polarized or inter-valley coherent, but never valley polarized~\cite{zhou2021half,arp2024intervalley,patterson2025superconductivity,auerbach2025visualizing}. 
Additionally, our argument suggests that states that preserve a momentum-negating symmetry gain more correlation energy; this may explain why half and full metals are favored over the three quarter metal. 
Our theory predicts that a $C_3$ symmetric system with valleys centered at the $M$ points will have a strong tendency towards valley polarization.

We thank Chunli Huang, Yves Kwan, and Yaar Vituri for useful discussions. 
V.C., S.A.K and E.B. were funded, in part, by NSF-BSF award DMR-2310312. 
E.B. was supported by the European Research Council (ERC) under the European Union’s Horizon 2020 research and innovation programme (grant agreement No 817799) and by a Simons Foundation Collaboration on New Frontiers in Superconductivity (Grant SFI-MPS-NFS-00006741-03). 
E.B. and V.C. also acknowledge support by grant NSF PHY-2309135 to the Kavli Institute for Theoretical Physics (KITP).

\bibliography{References}

\newpage
\section{End Matter}

\subsection{
Lifting  degeneracies of the SP states in Case II}\label{app:PerturbationsSP}

{
In the main text, we have neglected various interactions on the grounds that (under appropriate circumstances) they are relatively weak. However, since the symmetry of the Hamiltonian is reduced in the presence of such couplings, even when they are weak, they can play a central roll in lifting degeneracies. 

The independent SU(2) spin rotational symmetries of each valley is lifted by 
(so far neglected) large momentum transfer pieces of the Coulomb interaction. These terms are smaller than those we have treated by a factor of order $\sqrt{\nel a^2} \sim k_Fa\ll 1$ where $a$ is the lattice constant (see e.g.~\cite{chatterjee2022inter}). 
These terms produce a ferromagnetic inter-valley Hund's coupling, i.e. in any state in which two (or more) valleys are spin-polarized, the aligned  (fully polarized spin ferromagnetic) state  is favored.

Ising SOC ($ \sigma^z \tau^z $) favors out-of-plane ($z$-direction) spin alignment, with opposite spin orientations in the two inequivalent valleys. This leads to an Ising anti-ferromagnetic (AFM) SP state with spins pointing in the $z$-direction. This term preserves a $\U(1)^{\times 4}$ continuous symmetry, corresponding to conservation of particles of each flavor. Note that this Ising-AFM SP does not spontaneously break any continuous symmetry.  Thus, the corresponding ordered phases should persist up to a finite transition temperature.

The Rashba SOC $(k_x\sigma^y-k_y \sigma^x  )$ breaks the spin rotation symmetry.  Assuming there exists an unbroken vertical mirror plane, this leads to a locking of the spin moment to the XY-plane.  Trigonal warping further leads to orienting the spin moment along appropriate crystallographic directions.  Moreover, this induces a corresponding deformation of
the Fermi surface.  

To understand the spin ordering in the $C_3$ invariant case, we consider the following effective Hamiltonian which has the leading order (in powers of $\vec S$) small terms of various allowed symmetries 
\begin{equation}
\begin{split}
    H_{\rm{eff}} = - J&\vec{S}_{+} \cdot \vec{S}_{-} 
    -\lambda (S^z_+-S^z_- )\\
    \quad & -\alpha( [S_+^x+\ii S_+^y]^3-[S_-^x+\ii S_-^y]^3 +{\rm{c.c.}}),
\end{split}
\end{equation}
where $\vec{S}_{\tau}=[S^x_{\tau},S^y_{\tau},S^z_{\tau}]$ is a (classical) vector representing the spin-polarization for electrons in valley $\tau = \pm $, $J>0$ is the effective Hund's coupling, $\lambda$ is the effective Ising SOC coupling, and $\alpha$ is the effective Rashba SOC couplings. In the regime $J \gg \abs{\alpha}, \abs{\lambda}$ the spins are mostly ferromagnetically  aligned in plane. 
We thus parametrize the spins as $\vec{S}_\tau = [
\cos(\theta)\cos(\beta +\tau \phi),
\cos(\theta)\sin(\beta+\tau \phi),
\tau \sin(\theta)]$, where $\theta$ and $\phi$ can be assumed to be small. 
The ground-state obtained by minimizing this expression has a discrete (6-fold) degeneracy, labeled by integers $0\leq m < 6$:
\begin{equation}
    \begin{split}
        \theta_\star \approx \frac{\lambda}{2J};\quad 
    \phi_\star \approx -\frac{3\alpha}{2J}(-1)^m; \\
    \beta_\star = \frac{(2m+1)\pi}{6}.
    \end{split}
\end{equation}
This is a rather subtle form of coexisting in-plane ferromagnetism and non-coplanar antiferromagnetism.  Clearly, other forms of order are possible in other limits.  For instance there is a unique (non symmetry broken) state if $|\lambda| \gg |J|$ and $|\alpha|$.

\subsection{Three valleys related by \texorpdfstring{$C_3$}{C3}}\label{app:EndMatter:ThreeValleys}

Here, we flesh out the details that lead to the predictions in Fig.~\ref{fig:fig_M_pts}. We denote the filling by a vector 
\begin{equation}
\vec{\nu}=(\nu_{1,\ua},\nu_{1,\da},\nu_{2,\ua},\nu_{2,\da},\nu_{3,\ua},\nu_{3,\da}),
\end{equation}
subject to the constrain $\sum_{a=1}^6 \nu_a=1$. 
In parallel with the case of two valleys related by $C_4$ \cite{calvera2024theory,Nematic2DEG_VMC}, 
we consider the possible phases in which $n$ flavors are equally populated and $6-n$ are empty, with $n=1$,$\dots$, 5, or 6. 

The cases with $n=1, 5$ and $6$ there are no artificial degeneracies to be lifted. $n=2$ corresponds  to case II in Fig.~\ref{fig:1}. For $n=3$, there are two inequivalent configurations, $\vec{\nu}_{A} = (1/3,1/3,1/3,0,0,0)$ and $\vec{\nu}_{B} = (1/3,0,1/3,0,1/3,0)$. We now write $\Xsf_1=v( \Pi_{M_1} +\frac{1}{2}\Pi_{M_2})$ and $\Xsf_2= \mathcal{M}_{x}\Xsf_1 = v(\Pi_{M_3}+\frac{1}{2}\Pi_{M_2})$, so that $v\Pi_{A} =\Xsf_1 +\Xsf_2$ and $v\Pi_{B} = 2\Xsf_1 $. As each valley is invariant under momentum negation, from the argument in case II, we obtain that $E_{A}<E_{B}$. For case $n=4$, there are two inequivalent states, $\vec{\nu}_{\rm{A}} = (1/4,1/4,1/4,1/4,0,0)$ and $\vec{\nu}_{\rm{B}}= (1/4,1/4,1/4,0,1/4,0)$. Setting $\Xsf_1 = v(\Pi_{M_1}+\Pi_{M_2})$ and $\Xsf_2= C_3\Xsf_1 = v(\Pi_{M_2}+\Pi_{M_3})$, so that $v\Pi_{A} =2 \Xsf_1$ and $v\Pi_{B} = \Xsf_1 + \Xsf_2$, we see that $E_{A}<E_B$.

We expect that the 5F and 3F states are unstable to the transfer of electrons from or to the spin-polarized valley to the spin-degenerate valleys because the latter states do not break any further symmetry. 

\newpage 
\onecolumngrid

\section{Polarization bubbles}\label{app:BubbleProps}

\subsection{Bare polarization bubble}\label{app:BubbleProps:bare}
From the explicit formula in Eq.~\ref{eq:bare_bubble}, 
\begin{gather}
    [\Pi_{\alpha}(\qbs,\ii\omega)]^* = \Pi_{\alpha}(\qbs,-\ii\omega)\\
    \Pi_{\alpha}(-\qbs,-\ii\omega)= \Pi_{\alpha}(\qbs,\ii\omega)
\end{gather}
In other words, the real and imaginary parts of $\Pi_{\alpha}(\qbs,\ii\omega)$ are even and odd, respectively under $\qbs \to -\qbs$ and $\omega \to -\omega$. As $\Pi$ is calculated using occupations numbers and the bare energies $n_{\kbs,\alpha} = f(\ep_{\kbs,\alpha})$, where $f(\ep)$ is a decreasing function, which at zero temperature is a step function $\theta(E_{F,\alpha}-\ep)$. At finite temperature, it is the Fermi occupation factor $\frac{1}{e^{\beta(\ep-\mu)}+1}$. The real part satisfies
\begin{equation}
    \Re\Pi_{\alpha}(\qbs,\ii\omega) = \frac{1}{\Area}\sum_{\kbs} \frac{-(\ep_{\kbs+\qbs,\alpha}-\ep_{\kbs,\alpha})(n_{\kbs+\qbs,\alpha}-n_{\kbs,\alpha})}{\omega^2+(\ep_{\kbs+\qbs,\alpha}-\ep_{\kbs,\alpha})^2}\abs{\Lambda_{\kbs+\qbs,\kbs}^{\alpha}}^2 \geq 0.
\end{equation}
This follows from the observation that $f(x)-f(y)$ has the same sign as $-(x-y)$ when $f$ is a decreasing function. This shows that the numerator in the summand is positive. As all other terms are clearly positive, the inequality follows. 

There is no definite sign on the imaginary part of the polarization function:
\begin{equation}
    \Im\Pi_{\alpha}(\qbs,\ii\omega) = \frac{1}{\Area}\sum_{\kbs} \frac{-(\omega)(n_{\kbs+\qbs,\alpha}-n_{\kbs,\alpha})}{\omega^2+(\ep_{\kbs+\qbs,\alpha}-\ep_{\kbs,\alpha})^2}\abs{\Lambda_{\kbs+\qbs,\kbs}^{\alpha}}^2.
\end{equation}

We additionally see that for a state such that there is some symmetry $\gel$ such that the energies, occupations numbers and matrix elements for $\alpha$ and $\gel(\alpha)$ related by $\abs{\Lambda_{\kbs,\kbs'}^{\qbs;\alpha}}= \abs{\Lambda_{\gel(\kbs),\gel(\kbs')}^{\gel(\qbs);\gel(\alpha)}}$, give
\begin{equation}
    \Pi_{\alpha}(\qbs,\ii\omega) = \Pi_{\gel(\alpha)}(\gel(\qbs),\ii\omega).
\end{equation}
For example, if one flavor is symmetric under $\qbs\to -\qbs$, then the above shows that the polarization bubble is real. 

An important case is when two valleys $\alpha$ and $\bar{\alpha}$ are related by time-reversal or spatial inversion. Both symmetries act on momentum as $\qbs\to-\qbs$. Then,
\begin{equation}
    \Pi_{\bar{\alpha}}(\qbs,\ii \omega) = \Pi_{{\alpha}}(-\qbs,\ii \omega) = \Re\Pi_{\alpha}(\qbs,\ii\omega) - \ii \Im\Pi_{\alpha}(\qbs,\ii\omega).
\end{equation}
In particular, $\Pi_{\alpha} + \Pi_{\bar{\alpha}} = 2\Re\Pi$ so that $2\Pi_{\alpha}$ has the same real part as $\Pi_{\alpha}+\Pi_{\bar{\alpha}}$ but generically a non-zero imaginary part. 

\subsection{Plasmon dispersion}\label{app:Plasmon}
\def\PP{\mathbb{P}}
To find the plasmon dispersion in the RPA, we need to look for poles in the susceptibility
\begin{equation}
\chi(\qbs,z)=\frac{\chi_0(\qbs,z)}{1+v_{\qbs}\chi_0(\qbs,z)},
\end{equation}
where $z$ is a complex variable in the upper-half plane and $\chi_0$ is the analytical continuation of the bare susceptibility. 

The pole can be obtained by solving 
\begin{equation}\label{eq:PlasmonEq}
    1 + v_{\qbs} \chi_0(\qbs,z) =0.
\end{equation}
Assuming that $v_{\qbs}$ is large, we can look for solutions by doing a large $z$ expansion. To do so, we study the polarization bubbles 
\begin{equation}
    \Pi_{\alpha}(\qbs,z)  = \sum_{n=1}^{\infty} \frac{-\varpi_{\alpha;n}(\qbs)}{z^n}
\end{equation}
where $\rho_0 = \lim_{\qbs \to \zero} \Pi(\qbs,0)$ is the density of states and 
\begin{gather}
    \varpi_{\alpha;n}(\qbs) = \frac{1}{\Area}\sum_{\kbs} (\ep_{\kbs+\qbs,\alpha}-\ep_{\kbs,\alpha})^{n}\frac{-(n_{\kbs+\qbs,\alpha}-n_{\kbs,\alpha}) }{(\ep_{\kbs+\qbs,\alpha}-\ep_{\kbs,\alpha}) }\abs{\Lambda_{\kbs+\qbs,\kbs}^{\alpha}}^2.
\end{gather}
Note that $\varpi_{\alpha;n}(-\qbs) = (-1)^n\varphi_{\alpha;n}(\qbs)$. In the limit $\qbs\to \zero$, $\varpi_{\alpha;n}(\qbs)$ is dominated by the $\kbs$ near the Fermi surface, so we can can approximate $(\ep_{\kbs+\qbs,\alpha}-\ep_{\kbs,\alpha}) \approx \qbs \cdot \vbs_{F}(\kbs)$. In this regime,
\begin{equation}
    \varpi_{\alpha;n}(\qbs)  \approx \int_{\kbs} \delta(\ep_{\kbs,\alpha} - \mu_{\alpha}) (\qbs \cdot \vbs_{F}(\kbs))^n; \quad (\qbs \to \zero ),
\end{equation}
where $\mu_{\alpha}$ is the chemical potential of flavor $\alpha$. In other words, $\varpi_{\alpha;n}(\qbs)$ is roughly the integral of the product of the Fermi velocity and $\qbs$ around the Fermi surface. 

As $\chi_0(\qbs,z) = \sum_{\alpha}\Pi_{\alpha}(\qbs,z)$, there is a similar expansion 
\begin{equation}
    \chi_0(\qbs,z) = \sum_{n=1}^{\infty} -\frac{\Xmc_n(\qbs)}{z^n},
\end{equation}
where $\Xmc_n = \sum_{\alpha}\varpi_{\alpha;n}$. We can use this expansion solve Eq.~\ref{eq:PlasmonEq} order-by-order in the small parameter $\gamma_{\qbs}= 1/\sqrt{\rho_0v_{\qbs}}$ where $\rho_0 = \lim_{\qbs\to \zero}\chi_0(\qbs,0)$ is the density of states. The expansion for the energy is
\begin{equation}
    E_{\rm{pl}}(\qbs) = \omega_0(\qbs) \left(1+ \sum_{n=1}^{\infty}\gamma_{\qbs}^{n} \eta_{n}(\qbs)\right),
\end{equation}
with $\omega_0 = \sqrt{\Xmc_2(\qbs)}/\gamma_{\qbs}$. This expansion is controlled when $\Xmc_1(\qbs)/\gamma_{\qbs}^2\sqrt{\Xmc_2(\qbs)} \ll 1$, which holds for Coulomb-like interactions in the presence of $C_3$ or $C_2$ rotation symmetry.  

We leave the justification of the expansion to Sec.~\ref{app:solvingPlasmon}. Instead, we proceed to look at trends of the plasmon energies on each case. 

\textit{Case I.--} At 
order in $\gamma_{\qbs}$, the plasmon dispersion already distinguishes between spin-polarized (SP) and valley-polarized (VP) states. For $\qbs$ small,
\begin{equation}
    E_{\rm{pl}}^{\SP} = \sqrt{\omega_{0,1}^2 + \omega_{0,3}^{2}}, \quad 
    E_{\rm{pl}}^{\VP} = \sqrt{\omega_{0,1}^2 + \omega_{0,2}^{2}} = 2\omega_{0,1},
\end{equation}
where $\omega_{0,\alpha}$ denotes the leading-order plasmon frequency for flavor $\alpha$. Since $\sqrt{x}$ is concave, it follows that
\begin{equation}
    E_{\rm{pl}}^{\VP}(\qbs) + E_{\rm{pl}}^{\VP}(\gel\qbs) 
    \leq E_{\rm{pl}}^{\SP}(\qbs) + E_{\rm{pl}}^{\SP}(\gel[\qbs]),
\end{equation}
where $\gel$ is the symmetry exchanging flavors $1$ and $3$. Thus, VP states have a lower average plasmon energy than SP states, consistent with the energetic preference found in the correlation energy.

\textit{Case II.--} Here, the leading-order term $\omega_0(\qbs)$ is the same for SP and VP states because $\Xmc_2(\qbs)$ is even in $\qbs$. The distinction emerges at first order in $\gamma_{\qbs}$ via the $\eta_1(\qbs)$ term. Symmetry ensures that $\eta_1$ vanishes for SP, while it is generally nonzero for VP. However, since $\eta_1(\qbs)$ is odd, it contributes with opposite signs at $+\qbs$ and $-\qbs$, making the average plasmon energy {($\bar{E}_{\rm{pl}}(\qbs) = \frac{{E}_{\rm{pl}}(\qbs)+{E}_{\rm{pl}}(-\qbs)}{2}$) }
the same to this order for SP and VP. The first nontrivial distinction in average energy appears at second order through $\eta_2$, which we compute in App.~\ref{app:solvingPlasmon}. Interestingly, we find that $\eta_2$ is generically non-negative, and---neglecting form factors---VP plasmons have lower average energy than SP plasmons. This implies that the energetic preference for the SP state in Case II arises not from the plasmon alone, but from the full set of collective modes.

\subsubsection{Bounds of expansion coefficients}
The following interpretation for the coefficients $\varpi_{\alpha;n}(\qbs)$ is useful: $\varpi_{\alpha;n}(\qbs)$ is the expectation value of the variable $U_{\alpha,\qbs}(\kbs) \equiv (\varepsilon_{\kbs+\qbs,\alpha}- \varepsilon_{\kbs,\alpha})$ with probability $\PP_{\qbs,\alpha}(\kbs) =  \frac{1}{\omega_{0;\alpha}(\qbs)}\frac{1}{A}\frac{-(n_{\kbs+\qbs,\alpha}-n_{\kbs,\alpha}) }{(\ep_{\kbs+\qbs,\alpha}-\ep_{\kbs,\alpha}) }\abs{\Lambda_{\kbs+\qbs,\kbs}^{\alpha}}^2 $, as long as $\varpi_{0;\alpha}(\qbs) > 0 $. $\varpi_{0;\alpha}(\qbs) =0$ only when flavor $\alpha$ is unoccupied so we can just omit $\Pi_{\alpha}$ in the calculation of $\chi_0$. This also let us conclude that the even moments are non-negative $\varpi_{\alpha,2n}(\qbs)\geq 0$.

Note also that $\varpi_{\alpha;1}(\qbs) = \frac{1}{\Area}\sum_{\kbs}n_{\kbs,\alpha}\left(\abs{\Lambda_{\kbs+\qbs,\kbs}^{\alpha}}^2-\abs{\Lambda_{\kbs-\qbs,\kbs}^{\alpha}}^2\right)$. Therefore, if $\Lambda^{\alpha}=1$ then $\varpi_{\alpha,0}$ vanishes. On the other hand, when $\qbs$ is small $\varpi_{\alpha;1}(\qbs)$ is roughly the average of the Fermi velocity along the Fermi surface. Therefore, if flavor $\alpha$ is invariant under $C_2$ or $C_3$. Actually for $C_2$ invariant case $\varpi_{\alpha;1}$ must vanish because it is an odd function of $\qbs$. For the $C_3$ invariant case, $\varpi_{\alpha;1}$ will go at least as $\qbs \times \abs{\qbs}^2$.

\subsubsection{Solving for the plasmon pole}\label{app:solvingPlasmon}

We can rewrite Eq.~\ref{eq:PlasmonEq} in terms of the large $z$ expansion as 
\begin{equation}
    \gamma_{\qbs}^2 = \sum_{n=1}^{\infty} \frac{\Xmc_n(\qbs)}{z^n},
\end{equation}
with $\gamma_{\qbs} = 1/\sqrt{\rho_0 v_{\qbs}}$ is a small parameter in our expansion (for the spin-polarized 2DEG $\gamma_{\qbs} = \sqrt{a_{\rm{B}} q}$). As $\Xmc_1(\qbs)$ is going to be small, we can start by looking at the quadratic term $\gamma_{\qbs}^2 = \frac{\Xmc_2(\qbs)}{z^2}$ to obtain $z = \frac{1}{g_{\qbs }}\sqrt{\Xmc_2(\qbs)} =: \omega_{0}(\qbs)$. Making a change of variables to $\xi = \frac{z}{\omega_0(\qbs)}$ gives us
\begin{equation}
    1 = \frac{\Xmc_1(\qbs)}{\gamma_{\qbs}\sqrt{\Xmc_2(\qbs)}}\frac{1}{\xi} + \frac{1}{\xi^2} + \sum_{n=3}^{\infty} \gamma_{\qbs}^{n-2} \frac{\Xmc_{n}(\qbs)}{\left[\Xmc_2(\qbs)\right]^{n/2}}\frac{1}{\xi^n}.
\end{equation}
For small $\qbs$, $\Cmc_n(\qbs)\equiv\Xmc_n(\qbs)/[\Xmc_2(\qbs)]^{n/2}$ is in general an order one number because the powers of $\abs{\qbs}$ cancel. Assuming that $\Amc_{\qbs} \equiv \frac{\Xmc_1(\qbs)}{\gamma_{\qbs}^2 \sqrt{\Xmc_2(\qbs)}}$ is not too big (which is true in the small $\qbs$ regime in the presence of $C_3$ or $C_2$ symmetry), we can find a perturbative solution in powers of $\gamma_{\qbs}$, by setting $\xi= 1+ \sum_{n=1}^{\infty} \gamma_{\qbs}^n \eta_n(\qbs)$ and comparing term by term. To second order in $\gamma_{\qbs}$ we obtain
\begin{equation}
    \frac{z}{\omega_0(\qbs)} = 1 - \frac{\Amc_{\qbs}+\Cmc_3(\qbs) }{2}\gamma_{\qbs} + \frac{4(\Cmc_4(\qbs)-\Cmc_3(\qbs)^2) + (\Amc_{\qbs}-\Cmc_3(\qbs))^2}{8}\gamma_{\qbs}^2 + \mathcal{O}(g_\qbs^3).
\end{equation}
We can then read off $\eta_1(\qbs) = -\frac{\Amc_{\qbs}+\Cmc_3(\qbs)}{2}$ and $\eta_2(\qbs) = \frac{4(\Cmc_4(\qbs)-\Cmc_3(\qbs)^2) + (\Amc_{\qbs}-\Cmc_3(\qbs))^2}{8}\gamma_{\qbs}^2$. We proceed to show that $\eta_2(\qbs) \geq 0$.

The Cauchy-Schwartz inequality can be used to prove that
\begin{equation}
    \left[\int X^{p+q} \dd\upmu \right]^2 \leq 
    \left[\int X^{2p} \dd\upmu\right] 
    \left[\int X^{2q} \dd\upmu\right] ,
\end{equation}  
for any measure $\upmu$, real function $X$ and integers $p$ and $q$. Using the measure $\dd{\upmu}: (\kbs,\alpha) \to \varpi_{\alpha;0}(\qbs)\PP_{\qbs,\alpha}(\kbs)$ and variable $X:(\kbs,\alpha)\to \ep_{\kbs+\qbs,\alpha} -\ep_{\kbs, \alpha}$, gives
\begin{equation}
    [\Xmc_{p+q}(\qbs)]^2 \leq \Xmc_{2p}(\qbs) \Xmc_{2q}(\qbs).
\end{equation}
Taking $p=2$ and $q=1$ and dividing by $\Xmc_2(\qbs)^2$, gives us
\begin{equation}
    \Cmc_3(\qbs)^2\leq \Cmc_4(\qbs).
\end{equation}
As $\eta_2$ is $\frac{\Cmc_4-\Cmc_3^2}{2}$ plus a square, we conclude that $\eta_2 \geq 0$.

\subsubsection{Hartree-Fock bubble}\label{app:TDHF_chi}

\begin{figure}[t]
    \centering
\includegraphics[width=0.75\linewidth]{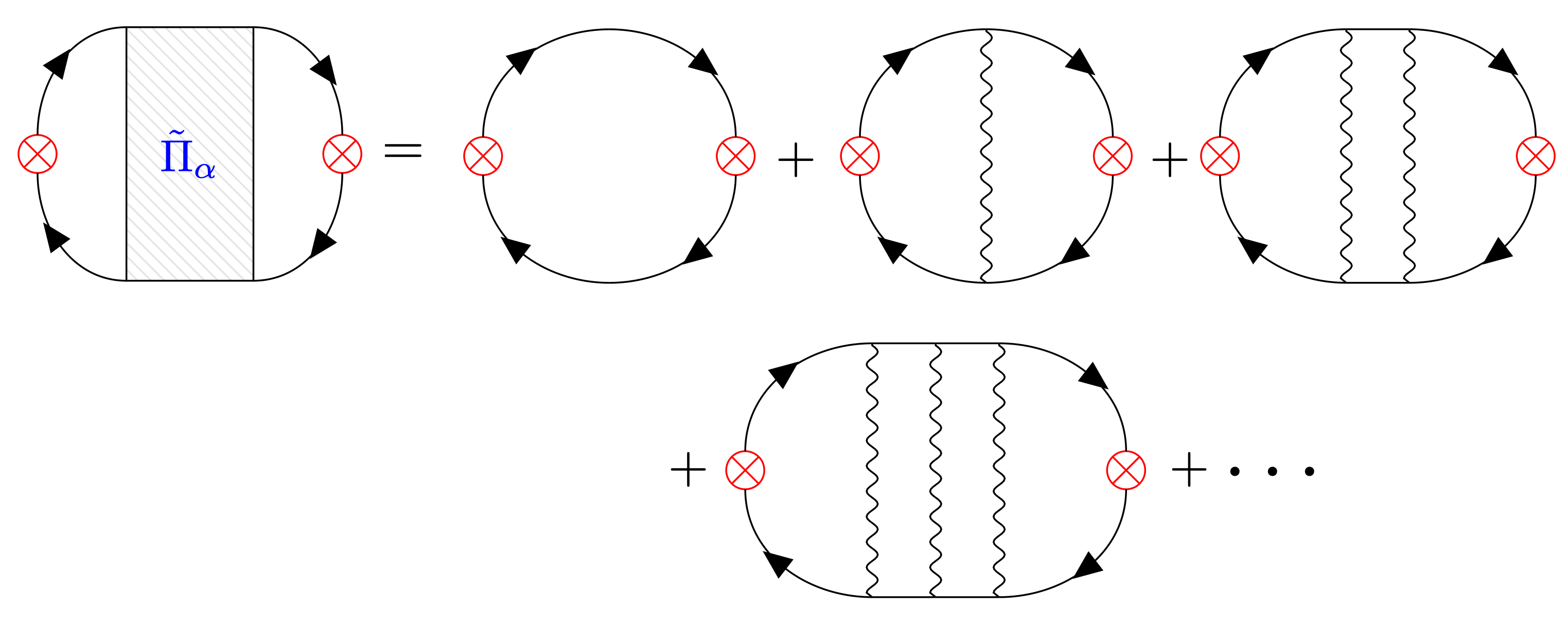}    \caption{The renormalized polarization bubble ($\tilde{\Pi}_{\alpha}$) is obtained by dressing the bare bubble by interactions. Black lines denote fermion Green's functions including the Hartree-Fock self energy and include the Fock self-energy. Wavy lines denote bare interactions. Red dots denote insertion of density operators. }
    \label{fig:fig2}
\end{figure}
To calculate the charge susceptibility in time-dependent Hartre-Fock (TDHF) one needs to include exchange diagrams in addition to the direct diagrams included in RPA. This is equivalent to including vertex corrections on the irreducible charge susceptiblity, {i.e.} one needs to compute the polarization bubble including exchange as shown in Fig.~\ref{fig:fig2}:
\begin{equation}
    \tilde{\Pi}_{\alpha}(\qbs,\ii\omega)
    -{\Pi}_{\alpha}(\qbs,\ii\omega) = \sum_{n=1}^{\infty} {\Pi}_{\alpha;n}(\qbs,\ii\omega),
\end{equation}
where $\tilde{\Pi}_{\alpha;n}(\qbs,\ii\omega)$ is the contribution with $n$ interaction lines:
\begin{equation}\label{eq:TDHF:correction:n}
    {\Pi}_{\alpha;n}(\qbs,\ii\omega) = (-1)^n\int_{\{\omega_j,\pbs_j\}_{j=0}^n}\prod_{j=1}^{n} v_{\pbs_{j-1}-\pbs_j}\Tr[
    \prod_{j=0}^{n}\Gmc_{\alpha}(\ii \omega_j,\pbs_j)
    \prod_{j=0}^{n}\Gmc_{\alpha}(\ii \omega_{n-j}+\ii\omega,\pbs_{n-j}+\qbs)],
\end{equation}
where $\Gmc_{\alpha}(\kbs,\ii\omega) = \frac{\ket{\kbs,\alpha}\bra{\kbs,\alpha}}{\ii\omega-(\bar{\varepsilon}_{\kbs,\alpha}-E_{F,\alpha})}$, $\prod_{j=0}^n O_j \equiv O_0 \cdot O_1 \cdots O_n $, $\int_{\{\omega_j,\pbs_j\}_{j=0}^n} \equiv \int_{\omega_0,\pbs_0}\int_{\omega_1,\pbs_1}\cdots \int_{\omega_n,\pbs_n}$ and $\bar{\varepsilon}_{\kbs}$ is the dispersion including the Fock self-energy.

The change of variables $\omega_j \to \omega_{n-j}-\omega$ and $\pbs_{j}\to \pbs_{n-j}-\qbs$ shows that ${\Pi}_{\alpha;n}(\qbs,\ii\omega) = {\Pi}_{\alpha;n}(-\qbs,-\ii\omega)$. Using the fact that $\Tr[O]^*=\Tr[O^{\dagger}]$ and that $\Gmc(\ii\omega,\pbs)^\dagger = \Gmc(-\ii\omega,\pbs)$ shows that ${\Pi}_{\alpha;n}(\qbs,\ii\omega)^* = {\Pi}_{\alpha;n}(-\qbs,\ii\omega)$ after a change of variables $\omega_j \to -\omega_{n-j}-\omega$ and $\pbs_{j}\to \pbs_{n-j}-\qbs$.

If there is a unitary $U_g$ such that $U_g \ket{\kbs,\alpha} = \ket{g[\kbs],g[\alpha]}$ and $E_{F,\alpha}=E_{F,g[\alpha]}$, we can insert $U_{g}U_g^\dagger$ between each $\Gmc$ in Eq.~\ref{eq:TDHF:correction:n} to obtain 
\begin{equation}
    \Pi_{\alpha;n}(\qbs,\ii\omega) 
    =\Pi_{g[\alpha];n}(g[\qbs],\ii\omega).
\end{equation}
If instead $U_g \ket{\kbs,\alpha}^* = \ket{g[\kbs],g[\alpha]}$, we have 
\begin{equation}
\Pi_{\alpha;n}(\qbs,\ii\omega)=    \Pi_{\alpha;n}(-\qbs,\ii\omega) ^* = 
    \Pi_{g[\alpha];n}(-g[\qbs],-\ii\omega)  = \Pi_{g[\alpha];n}(g[\qbs],\ii\omega),
\end{equation}
where we used that $\Tr[O] =\Tr[O^*]$ together with $\Gmc_{\alpha}(\ii\omega,\qbs) = U_{g}^{\dagger}\Gmc_{g[\alpha]}(-\ii\omega,\qbs) U_g$.

As each $\Pi_{\alpha;n}$ satisfies the properties (1) and (2) of $\Pi_{\alpha}$ discussed in the main text, so will $\tilde{\Pi}_{\alpha}$. 

The TDHF charge susceptibility is 
\begin{gather}
    \chi(\qbs,\ii\omega) = \frac{\chi_1(\qbs,\ii\omega)}{1+v_{\qbs}\chi_1(\qbs,\ii\omega)} \\
    \chi_1(\qbs,\ii\omega) = \sum_{\alpha}\tilde{\Pi}_{\alpha}(\qbs,\ii\omega).
\end{gather}

We also would like to point that for a model without form factors $\tilde{\Pi}_\alpha(\qbs,0)$ is a sum of positive contributions. To see this, we can evaluate the frequency integrals in $\Pi_{\alpha;n}$:
\begin{equation}
  \Pi_{\alpha;n}(\qbs,0)=  \int_{\{\pbs_j\}_{j=0}^n}\prod_{j=1}^{n} v_{\pbs_{j-1}-\pbs_j}\prod_{j=0}^n \frac{n_{\pbs_j+\qbs}-n_{\pbs_j+\qbs}}{\tilde{\varepsilon}_{\pbs_j+\qbs}-\tilde{\varepsilon}_{\pbs_j}},
\end{equation}
which means that, in this situation,  $\chi(\qbs,0)$ cannot diverge if $v_{\qbs}$ is finite.

\section{Second Order Perturbation theory}\label{app:2PT}

\begin{figure}[t]
    \centering
    \includegraphics[width=0.85\linewidth]{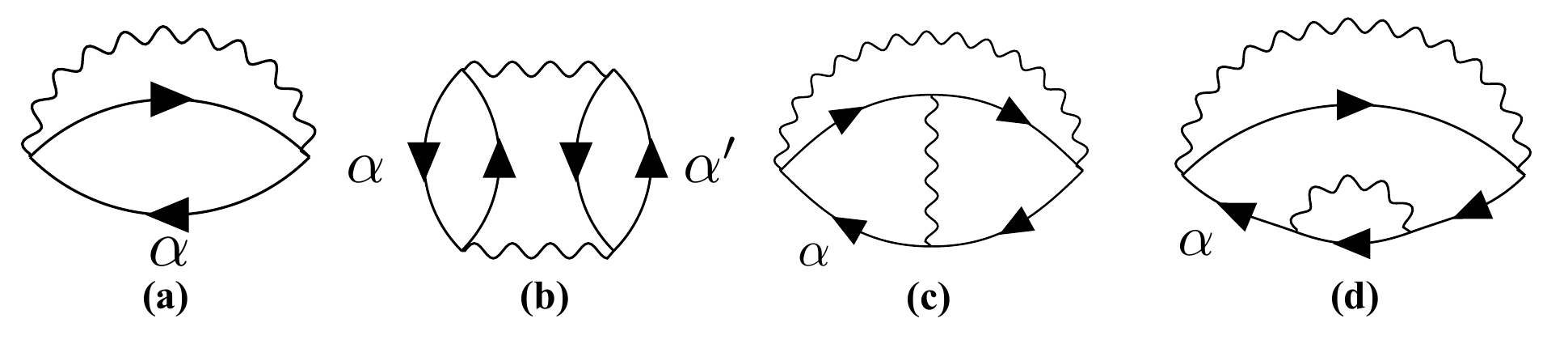}
    \caption{Diagrams contributing the energy up to second order perturbation theory. We have omitted the Hartree diagram.}
    \label{fig:2PT}
\end{figure}

To calculate the energy (in a polarization sector) to second order in interaction, we have to include the diagrams in Fig.~\ref{fig:2PT}. The Hartree diagram vanishes due to the neutralizing background. Diagram (a) is the Fock diagram

The second contribution (diagram b) gives ($E_{\rm{r}}=\sum_{\alpha,\alpha'}E_{\rm{r}}(\alpha,\alpha')$):
\begin{equation}
    E_{\rm{r}}(\alpha,\alpha') = -\frac{1}{4
    }\sum_{\qbs} v_{\qbs}^2\int_{-\infty}^{\infty} \Pi_{\alpha}(\qbs,\ii\omega)
    \Pi_{\alpha'}(\qbs,\ii\omega) \frac{\dd{\omega}}{2\pi},
\end{equation}
where the polarization bubbles are given by 
\begin{equation}
    \Pi_{\alpha}(\qbs,\ii\omega) = \int_{\kbs} \frac{n_{\kbs+\qbs}-n_{\kbs}}{\ii\omega -(\varepsilon_{\kbs+\qbs}^{\alpha}-\varepsilon_{\kbs}^{\alpha})}\abs{\Lambda_{\kbs,\kbs+\qbs}^{\alpha}}^2 = \int_{0}^{\infty}\left[\frac{\Asf_{\alpha}(\qbs,\varepsilon)}{\varepsilon-\ii\omega}  + \frac{\Asf_{\alpha}(-\qbs,\varepsilon)}{\varepsilon+\ii\omega} \right]\dd{\varepsilon},
\end{equation}
with $\Asf_{\alpha}(\qbs,\varepsilon)=\int_{\kbs}\abs{\Lambda_{\kbs,\kbs+\qbs}^{\alpha}} n_{\kbs}(1-n_{\kbs+\qbs}) \delta(\epsilon -[\varepsilon_{\kbs+\qbs}^{\alpha}-\varepsilon_{\kbs}^{\alpha}]) \geq 0$. Performing the frequency integration gives
\begin{equation}
    E_{\rm{r}}(\alpha,\alpha') = -\frac{1}{2}\sum_{\qbs}v_{\qbs}^2\int_{\varepsilon,\varepsilon'}
   \frac{ \Asf_{\alpha}(+\qbs,\varepsilon)
    \Asf_{\alpha'}(-\qbs,\varepsilon')}{\varepsilon+\varepsilon'}.
\end{equation}
We can express the result in terms of the Laplace transforms $\tilde{\Asf}_{\alpha}(\qbs,\tau)\equiv\int_0^{\infty}\Asf_{\alpha}(\qbs,\varepsilon)e^{-\varepsilon\tau}\dd{\varepsilon} $, so that  
\begin{equation}
    E_{\rm{r}}(\alpha,\alpha') = -\frac{1}{2}\sum_{\qbs}v_{\qbs}^2\int_{0}^{\infty}
   { \tilde{\Asf}_{\alpha}(+\qbs,\tau)
\tilde{\Asf}_{\alpha'}(-\qbs,\tau)} \dd{\tau}= - (\tilde{\Asf}_{\alpha},\mathsf{I}[\tilde{\Asf}_{\alpha'}]),
\end{equation}
where for real functions $f_1,f_2$ of momenta and $\tau$, $\mathsf{I}[f_1](\qbs,\tau)\equiv f_1(-\qbs,\tau)$ and $(f_1,f_2)\equiv \frac{1}{2}\sum_{\qbs}v_{\qbs}^2\int_0^{\infty}f_1(\qbs,\tau)f_2(\qbs,\tau) \dd{\tau}$. As $(.,.)$ is an inner product, we have the Cauchy-Schwartz inequality $\abs{(f_1,f_2)} \leq \sqrt{(f_1,f_1)(f_2,f_2)}$, with equality attained only when $f_1\propto f_2$. This means that $E_{\rm{r}}(\alpha,\alpha') \geq -\sqrt{
(\tilde{A}_{\alpha},\tilde{A}_{\alpha})
(\tilde{A}_{\alpha'},\tilde{A}_{\alpha'})
}$ with equality achieved only when $\tilde{A}_{\alpha'} = \mathsf{I}[\tilde{A}_{\alpha} ]$, which is equivalent to $\Asf_{\alpha}(\qbs,\varepsilon)=\Asf_{\alpha'}(-\qbs,\varepsilon)$. 

Diagram (c) gives a correction to the exchange given by 
\begin{equation}
    E_{\rm{x}}(\alpha) = \frac{1}{4
    }\sum_{\qbs} 
    \int_{\kbs,\kbs'}\int_{-\infty}^{\infty} \frac{n_{\kbs+\qbs}^{\alpha}-n_{\kbs}^{\alpha}}{\ii\omega -\varepsilon_{\kbs+\qbs}^{\alpha}+\varepsilon_{\kbs}^{\alpha}}
\frac{n_{\kbs'+\qbs}^{\alpha}-n_{\kbs'}^{\alpha}}{\ii\omega -\varepsilon_{\kbs'+\qbs}^{\alpha}+\varepsilon_{\kbs'}^{\alpha}}v_{\kbs-\kbs'}v_{\qbs}
{
\left[
\Lambda_{\kbs+\qbs,\kbs}^{\alpha}
\Lambda_{\kbs,\kbs'}^{\alpha}
\Lambda_{\kbs',\kbs'+\qbs}^{\alpha}
\Lambda_{\kbs'+\qbs,\kbs+\qbs}^{\alpha}
\right]
}
    \frac{\dd{\omega}}{2\pi},
\end{equation}
which does not distinguish betweent the two-half metals because there is no correlation between different flavors.

Diagram (d) vanishes at zero temperature. At finite temperature, however, it contributes to a shift in the chemical potential, which in turn affects the total energy \cite{kohn1960ground}. Importantly, this contribution does not include any correlations between different flavors so it does not distinguish between SP and VP.

Thus, in 2PT, the difference between SP and VP arises from $E_{\rm{r}}(\alpha,\alpha')$, where $\alpha$ and $\alpha'$ are the two occupied flavors for each half-metal. In other words the only term that makes a difference is 
\begin{equation}
    E_{\rm{r}}' = E_{\rm{r}}(1,\alpha')+E_{\rm{r}}(\alpha',1) = 2E_{\rm{r}}(1,\alpha'),
\end{equation}
with $\alpha' =2$ (3) for the VP (SP) sector. To obtain the expression quoted (Eq.~\ref{eq:2PT:ringPrimed}) in the main text, we use the property $\Pi_{\alpha}(\qbs,\ii\omega) = \Pi_{\alpha}(-\qbs,\ii\omega)^*$ combined with the fact that the integral must be real. The latter can be shown by making a change variables $\omega \to -\omega $ combined with $\Pi_{\alpha}(\qbs,\ii\omega)^*=\Pi_{\alpha}(\qbs,-\ii\omega)$.

As discussed above, this term is minimized when the electron-hole excitations with momenta $\qbs$ and flavor $\alpha$ (denoted $\Asf_{\alpha}(\qbs,\varepsilon)$) match those with momenta $-\qbs$ and flavor $\alpha'$ (denoted $\Asf_{\alpha'}(-\qbs,\varepsilon)$). In Case I, this condition is satisfied in the VP state, whereas in Case II, it is satisfied in the SP state. Note this argument is just restating what we said in the main text but working directly with the particle-hole spectra.

\section{Random phase approximation and coupling constant integration}\label{app:CouplingConstant:FH}

The aim of this appendix is to related the integration of coupling constant formula in Eq.~\ref{eq:CorrelationEnergy} of the main text to the use of the Feynman-Hellmann (FH) theorem combined with the time-dependent Hartree approximation for correlation functions (see e.g. \cite{Giuliani_Vignale_2005,barlas2007chirality} for a similar discussion). 

The HF theorem tells us that given a Hamiltonian $H_{\lambda} = H_0 + \lambda V$, the matrix element $\mel{\psi(\lambda)}{V}{\psi(\lambda)} = \pdv{E_{\psi}(\lambda)}{\lambda}$, where $\ket{\psi(\lambda)}$ is a first-differentiable family of eigenstates of $H_{\lambda}$ with energy $E_{\psi}$  ($H_{\lambda}\ket{\psi(\lambda)} = E_{\psi}(\lambda) \ket{\psi(\lambda)}$). We can then do an integration over $\lambda$ to obtain 
\begin{equation}\label{eq:Feynman-Hellman}
    E_{\psi}(\lambda=1) - 
    E_{\psi}(\lambda=0) 
    = \int_0^1 \mel{\psi(\lambda)}{V}{\psi(\lambda)}\dd{\lambda}.
\end{equation}
In other words, to calculate the energy of a given state, we need the unperturbed energy and the expectation value of the perturbation along the adiabatic path. 

We now specialize to the Hamiltonian in Eq.~\ref{eq:Ham} in the main text, with $H = H_0 + V$, where
\begin{equation}
        H_0  =\sum_{\kbs,\alpha}\ep_{\kbs,\alpha} c_{\kbs,\alpha}^{\dagger}c_{\kbs,\alpha}^{\,}; \quad
  V  = \frac{1}{2\Area} \sum_{\qbs \neq \mathbf{0}} v_{\qbs} \frac{:\rho_{\qbs}^{\dagger}\rho_{\qbs}+\rho_{\qbs}\rho_{\qbs}^{\dagger}:}{2}.
\end{equation}
As $H_0$ and $V$ commute with the number operators $N_\alpha = \sum_{\kbs} c^\dagger_{\kbs,\alpha}
c_{\kbs,\alpha}$, states with different $N_\alpha$ eigenvalues do not mix. Said it differently, we can project $H_0$ and $V$ to sectors with fixed values of $N_\alpha$. Therefore, assuming that the ground state in a given sector is connected to the interacting ground state, we can calculate the energy by studying the problem with just two flavors.

We can use a imaginary time version of the fluctuation-dissipation-theorem (see App.~\ref{app:FlucDissipation}) to show that the ground state expectation value is
\begin{equation}
    \frac{1}{2}\expval{\rho_{\qbs}^\dagger \rho_{\qbs}+ \rho_{\qbs}\rho_{\qbs}^\dagger} = \int_0^{\infty}\Re[\chi(\qbs,\ii\omega)]\frac{\dd{\omega}}{\pi},
\end{equation}
where $\chi(\qbs,\ii\omega)$ is the dynamic charge susceptibility. 

Note now that $:\rho^\dagger_{\qbs}\rho_{\qbs}:=\rho^\dagger_{\qbs}\rho_{\qbs} -\sum_{\alpha}\delta_{\qbs;\alpha} $ where $\delta_{\qbs;\alpha}\equiv \sum_{\kbs}\abs{\Lambda_{\kbs+\qbs,\kbs}^\alpha}^2c^\dagger_{\kbs+\qbs,\alpha}c_{\kbs+\qbs,\alpha}$. Therefore, 
\begin{equation}\label{eq:ExpvalV}
    2\expval{V} = \sum_{\qbs\neq \zero} \frac{v_{\qbs}}{\Area} \left[\int_0^{\infty} \Re[\chi(\qbs,\ii\omega)]\frac{\dd\omega}{\pi} - \expval{\sum_{\alpha}\delta_{\qbs;\alpha}}\right],
\end{equation}
which can be rewritten as $\expval{V}=\expval{V}_{\rm{x}}+\expval{V}_{\rm{c}}$, where
\begin{align}
    \expval{V}_{\rm{x}} = \sum_{\qbs\neq 0}\frac{v_{\qbs}}{2\Area} \left[\int_0^\infty \Re[\chi_0(\qbs,\ii\omega)]\frac{\dd\omega}{\pi}- \expval{\sum_{\alpha}\delta_{\qbs;\alpha}}\right]\\
    \expval{V}_{\rm{c}} = \sum_{\qbs\neq 0}\frac{v_{\qbs}}{2\Area} \left[\int_0^\infty \Re[\chi(\qbs,\ii\omega)-\chi_0(\qbs,\ii\omega)] \frac{\dd\omega}{\pi}\right]
\end{align}
In the absence of form factors, $\delta_{\qbs;\alpha}$ is the number operator, which is a conserved quantity. However, in our setup $\delta_{\qbs;\alpha}$ is not necessarily a conserved quantity. 

To obtain Eq.~\ref{eq:CorrelationEnergy} in the main text, we use the Feynman-Hellmann theorem (Eq.~\ref{eq:Feynman-Hellman}) combined with the time dependent Hartree approximation. As we have removed the $\qbs=\zero$ term, there is no Hartree self-energy so $\expval{\delta_{\qbs;\alpha}}$ is evaluated using the non-interacting Green's function. At coupling constant $\lambda$, $\chi(\qbs,\ii\omega)$ is approximated by
\begin{equation}
    \chi(\qbs,\ii\omega)_{\lambda} = \frac{\chi_0(\qbs,\ii\omega)}{1+\lambda v_{\qbs}\chi_0(\qbs,\ii\omega)}.
\end{equation}

\subsection{Fluctuation-dissipation theorem in imaginary time}\label{app:FlucDissipation}

Take two operators $A,B$ with zero ground-state expectation value $\mel{0}{A}{0} = \mel{0}{B}{0} =0$, where we denoted the ground state by $\ket{0}$. The susceptibility in Matsubara frequency is defined as
\begin{equation}
    \chi(A,B|\ii\omega ) : = \int_{-\infty}^{+\infty} \expval{\mathcal{T}[A(\tau)B]} e^{\ii\Omega \tau}\dd{\tau}; \quad \omega \in \mathbb{R},
\end{equation}
where $\mathcal{T}[A(\tau) B]$ is ordering and $A(\tau) =  e^{+\tau H} A e^{-\tau H}$, where we assumed the ground state has energy zero. The susceptibility satisfies the following properties: 1) $\chi(B,A|\ii\omega) = \chi(A,B|-\ii\omega)$; 2) $\chi(A,B|\ii\omega)^* = \chi(B^\dagger,A^\dagger, -\ii\omega)$; and 3) admits the following analytical continuation away from the real axis ($\ii\omega\to z$):
\begin{equation}\label{eq:chi:AnalyticalCont}
    \chi(A,B|z )  = \sum_{\ket{\ep}}^{'} \frac{\mel{0}{A}{\ep}\mel{\ep}{B}{0}}{\ep - z} + \frac{\mel{0}{B}{\ep}\mel{\ep}{A}{0}}{\ep + z};
\end{equation}
where the sum is over all the eigenstates of $H$ different than the ground state. 

Specializing to the case $B = A^\dagger$, we can write $\chi(A,B|\ii\omega)$ in terms of imaginary part of the retarded susceptibility:
\begin{equation}
    \begin{split}
        \chi(A,A^{\dagger}|z) &= \int^{\infty}_{-\infty} \frac{\chi{''}(A,A^\dagger|\omega) }{\omega - z}\frac{\dd{\omega}}{\pi} \\
        \chi{''}(A,A^\dagger|\omega)& =  \Im( \chi(A,A^\dagger|\omega+\ii0^+)) = 
        \pi{\sum_{\ket{\ep}}}^{'} \delta(\omega-\ep) \abs{\mel{0}{A}{\ep}}^2 -
          \delta(\omega+\ep) \abs{\mel{0}{A^\dagger}{\ep}}^2,\label{eq:chipp:SpectralExpansion}
    \end{split}
\end{equation}
where the sum is over excited states $\ket{\varepsilon}$.

The usual fluctuation-dissipation theorem, states that 
\begin{equation}
    \int^{+\infty}_{0} \chi{''}(A,A^\dagger|\omega) \frac{\dd{\omega}}{\pi} = \expval{AA^{\dagger}} \label{eq:Fluct:Diss}.
\end{equation}
This follows immediatly from Eq.~\ref{eq:chipp:SpectralExpansion}. We can rewrite the left hand side, as
\begin{equation}
    \int^{+\infty}_{0} \chi{''}(A,A^\dagger|\omega) \frac{\dd{\omega}}{\pi}  = \Im \int_{C} \chi(A,A^\dagger|z) \frac{\dd{z}}{\pi},
\end{equation}
where $C =[\ii 0^+,\infty+\ii 0^+)$ is the positive real axis infinitesimally displaced to the complex half-plane. 

From Eq.~\ref{eq:chi:AnalyticalCont}, one can see that $ \chi(A,A^\dagger|z) \frac{\dd{z}}{\pi} $ generically decays only as $1/z$ for $\abs{z}\to\infty$. This prevents us from deforming the contour $C$. To get around this, we study $\bar\chi(A|z)\equiv\frac{\chi(A,A^\dagger|z) + \chi(A^\dagger,A,|z)}{2}$:
\begin{equation}
  \bar\chi(A|z) =    {\sum_{\ket{\ep}}}^{'} \frac{\ep}{\ep^2 - z^2} \left(\abs{\mel{0}{A}{\ep}}^2 + \abs{\mel{0}{A^\dagger}{\ep}}^2\right).
\end{equation}
It is now clear that $\bar\chi(A|z)$ is analytic in the complex upper half plane that decays as $1/z^2$ for $\abs{z}\to\infty$. Using Eq.~\ref{eq:Fluct:Diss} we obtain 
\begin{equation}
    \Im \int_C \bar\chi(A|z) \frac{\dd{z}}{\pi} = \frac{1}{2}\expval{AA^\dagger+A^\dagger A}.
\end{equation}
As $ \bar\chi(A|z)$ decays at least as $1/z^2$, we can deform the contour $C$ to the imaginary axis without changing the integral:
\begin{equation}
    \Im \int_0^{\infty} \bar\chi(A|\ii\omega) \frac{\ii \dd{\omega}}{\pi} = \frac{1}{2}\expval{AA^\dagger+A^\dagger A}.
\end{equation}
Using property 1 and 2 of $\chi(A,B|\ii\omega)$, one can show that $\bar\chi(A|\ii\omega) = \Re[\chi(A,A^\dagger|\ii\omega)]$. Therefore, we can simplify
\begin{equation}
     \int_0^{\infty}  \Re[\chi(A,A^\dagger|\ii\omega)] \frac{ \dd{\omega}}{\pi} = \frac{1}{2}\expval{AA^\dagger+A^\dagger A}.
\end{equation}
In the main text, we use $A  = \rho_{\qbs}$ and, by definition, $\chi(\qbs,\ii\omega) \equiv \chi(\rho_{\qbs},\rho_{\qbs}^{\dagger}|\ii\omega)$.

\section{Extensions}\label{app:Extensions}

The aim of this appendix is to make more precise the the claim that the trends obtained in RPA do not change if we modify the RPA calculation. We briefly explain our claims regarding three dimensions and the orbital effect of the magnetic field below. We defer the explanation of partial polarization, multi-channel interaction, and time-dependent Hartree-Fock susceptibility to App.~\ref{app:PartialPolarization} and App.~\ref{app:MultiChannel}, respectively.

\textit{Three dimensions.--} Our arguments do not rely on the dimension of the system. For instance, consider a three dimensional electron system with a cubic BZ whose band minima are at the face centers. As these points are invariant under $D_4\times I$ (fourfold-rotations, reflection about a vertical plane and inversion), the longitudinal mass is generically different than the mass in the transverse directions. Repeating our argument, we find the same polarization tendencies as in Fig.~\ref{fig:fig_M_pts} if we interpret the $M$ points by the respective 3d momenta.

\textit{Orbital magnetic field.--} in the presence of an orbital magnetic field, momentum is no longer a quantum number. However, we can still diagonalize the one-particle Hamiltonian. We no longer can label eigenstates by momentum but that is not crucial. To see this, note that the Lindhard function is now of the form
\begin{equation}
    \Pi_{\alpha}(\qbs,\ii\omega) = \frac{1}{\Area}\sum_{i,j} \frac{n_i - n_j}{\ii\omega - (\varepsilon_{i} - \varepsilon_{j})} \abs{\Lambda_{i,j}^{\alpha}(\qbs)}^2,
\end{equation}
where $i$ and $j$ label an orthogonal basis of eigenstates of kinetic term, $n_i$ and $n_j$ their respective occupation, and $\Lambda_{i,j}^{\alpha}(\qbs) \equiv \mel{i,\alpha}{e^{\ii\qbs\cdot\rbs}}{j,\alpha}$ is the form factor. The orbital magnetic field explicitly breaks time reveral symmetry and mirror-symmetries but preserves the combination of those symmetries. Rotational symmetries present in the absence of the magnetic field are preserved albeit with a additional gauge transformation. The properties of $\Pi_{\alpha}$ remain true. For instance, $\Pi_{\alpha}(\qbs) = \Pi_{\gel[\alpha]}(\gel[\qbs])$ can be obtained by noting that orbitals occupied in state $\gel[\alpha]$ are obtained by acting on the occupied states in $\alpha$ by the symmetry $\gel$ so that we have $n_{\gel [i]}^{(\gel[\alpha])} = n_{i}^{(\alpha)}$ and $\abs{\Lambda^{\alpha}_{i,j}(\qbs)} = \Lambda^{\gel [\alpha]}_{\gel[i],\gel[j]}(g[\qbs])$. 

Our arguments should apply to integer fillings as the Slater determinants are good starting points. For instance, consider a situation where the lowest Landau level is only two-fold degenerate rather than four-fold. If the Zeeman effect is negligible, it is more likely that the degenerate LLs are valley polarized for Case I, and spin polarized for Case II. 

\textit{TDHF.--} if we calculated $\Smc$ using the TDHF susceptibility we would get the same trends. This is because the renormalized polarization bubbles $\tilde{\Pi}_{\alpha}$ introduced in App.~\ref{app:TDHF_chi} satisfy the same same properties as $\Pi_{\alpha}$ except for positivity of the real part. Therefore, as long as $\Re[v_{\qbs}\chi_1(\qbs,\ii\omega)]+1>0$, our arguments in the main text go throughout without any modification. However, this does not imply anything concrete for the correlation energy.

\subsection{Partial polarization}\label{app:PartialPolarization}

As mentioned in the main text, our results can be extended to the situation of partially polarized states. For concreteness, we compare the energies of a partially spin polarized state ($\rm{pSP}$) and a partially valley polarized state ($\rm{pVP}$). We assume that the density vector $(n_1,n_2,n_3,n_4)$ for $\pSP$ is $(\tfrac{\nel(1+\zeta)}{4},\tfrac{\nel(1-\zeta)}{4},\tfrac{\nel(1+\zeta)}{4},\tfrac{\nel(1-\zeta)}{4}
)$, and for $\pVP$ is $(\tfrac{\nel(1+\zeta)}{4},\tfrac{\nel(1+\zeta)}{4},\tfrac{\nel(1-\zeta)}{4},\tfrac{\nel(1-\zeta)}{4}
)$. where $- 1 \leq \zeta \leq 1$ is the degree of polarization. See Fig.~\ref{fig:partialPolarization} for a cartoon of the Fermi seas in Case II. 

\begin{figure}[t]
    \centering
\includegraphics[width=0.65\linewidth]{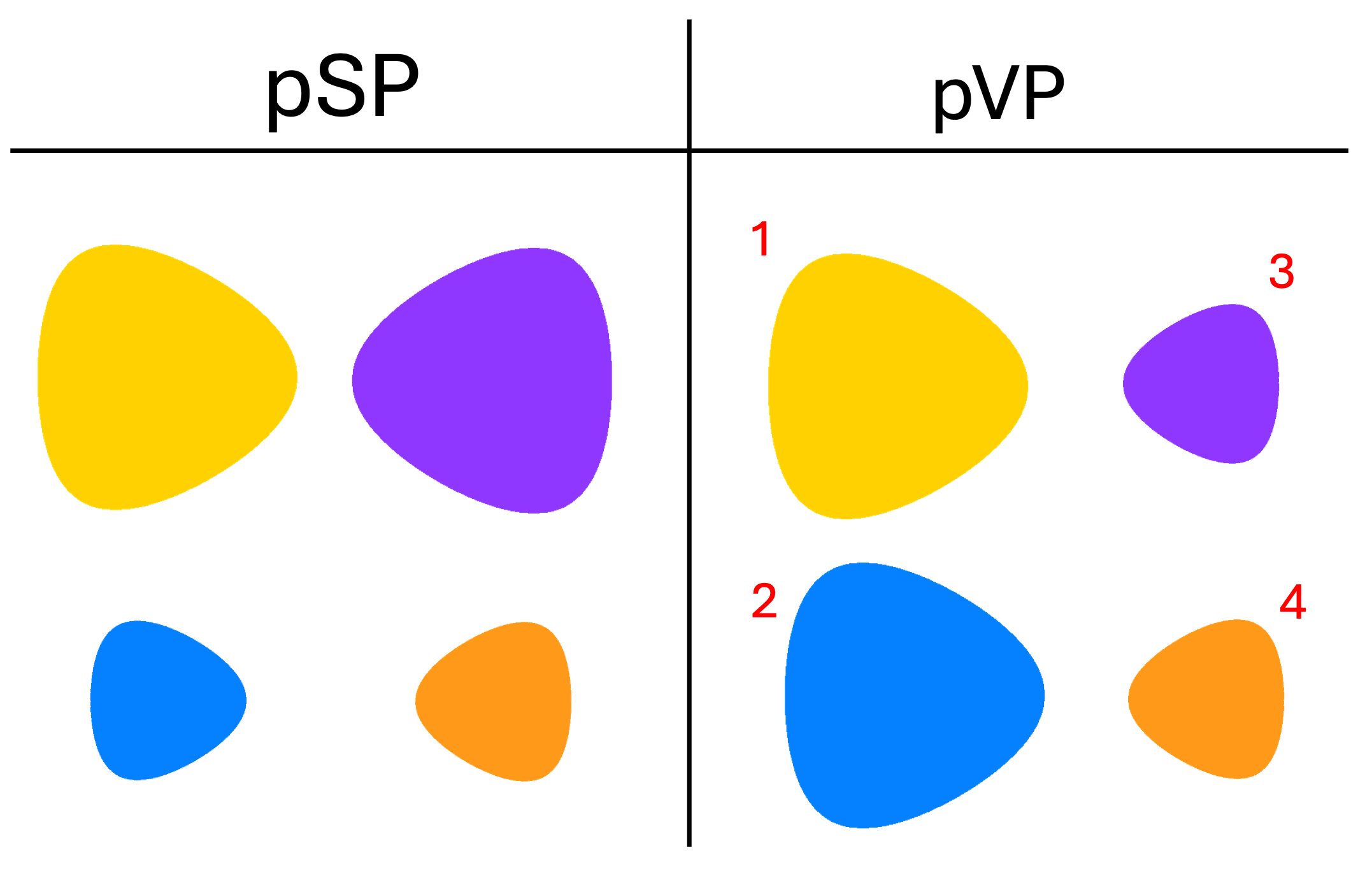}
    \caption{ Candidate states for partial half-metals in case II. pSP stands for partially spin polarized, and pVP stands for partially valley polarized.
    }
    \label{fig:partialPolarization}
\end{figure}

We find $\Pi_{\pSP} = \Pi_1^+ +  \Pi_1^- + \Pi_{3}^+ + \Pi_{3}^{-}$ and $\Pi_{\pVP} = 2\Pi_1^{+}+2\Pi_{3}^-$, where $\Pi_{\alpha}^{\pm}$ is shorthand for the polarization bubble of $\alpha$ electrons at density $\nel(1\pm \zeta)/4$. We reproduce the same trends in the correlation energy by repeating the arguments below Eq.~\ref{eq:StructureFactor} in the main text with $\Xsf_1 = v(\Pi_1^+ + \Pi_1^-)$ and $\Xsf_2 = v(\Pi_3^+ + \Pi_{3}^-)$ for Case I, and $\Xsf = v(\Pi_1^+ + \Pi_1^- + \Pi_{3}^+ + \Pi_3^-)$ and $\Ysf = v(\Pi_1^+ + \Pi_{1}^- - \Pi_{3}^- - \Pi_3^-)/\ii$. 

One way to calculate the spin and valley susceptibility in the symmetric state is by taking the limits 
\begin{equation}
    \begin{split}
    \frac{1}{\chi_{\rm{s}} } &\equiv \lim_{\zeta \to 0} \frac{E_{\pSP}(\zeta)-E_{\pSP}(0)}{\Area \frac{\zeta^2}{2}\nel^2},\\
    \frac{1}{\chi_{\rm{v}} } &\equiv \lim_{\zeta \to 0} \frac{E_{\pVP}(\zeta)-E_{\pVP}(0)}{\Area \frac{\zeta^2}{2}\nel^2},
    \end{split}
\end{equation}
where $E_{\pSP}(\zeta)$ and $E_{\pVP}(\zeta)$ are the energies of the pSP and pVP, respectively, at degree of polarization $\zeta$. 

Note that $E_{\pSP}(0)=E_{\pVP}(0) = E_{\rm{Sym}}$ so that we directly use our inequalities to show that 
\begin{equation}
\begin{split}
    \textbf{Case I:} &\quad\chi_{\rm{s}} \leq \chi_{\rm{v}}; \\
    \textbf{Case II:} &\quad\chi_{\rm{v}} \leq \chi_{\rm{s}}.
    \end{split}
\end{equation}

\subsection{Multiband and multichannel model}\label{app:MultiChannel}

Let's consider a more general problem where we include nearby bands and interactions can be more general than density-density interactions
\begin{gather}\label{eq:Ham:general}
    H = \sum_{\alpha=1}^{4} T_{\alpha} + V \\
    T_{\alpha}=\sum_{\kbs} c_{\kbs,\alpha}^{\dagger}h_{\kbs,\alpha} c_{\kbs,\alpha}^{\,}\\
    V =\frac{1}{2\Area} \sum_{\qbs,\mu,\nu}v_{\qbs}^{\mu\nu}:O_{\qbs,\mu}^\dagger O_{\qbs,\nu}:\\
    O_{\qbs,\nu} = \sum_{\alpha}O_{\qbs,\nu,\alpha}\\
    O_{\qbs,\nu,\alpha} = \sum_{\kbs} c_{\kbs+\qbs,\alpha}^{\dagger}\Upsilon_{\nu}c_{\kbs,\alpha}.
\end{gather}
Here $c_{\kbs,\alpha}^{\dagger} = [c_{\kbs,\alpha,1}^{\dagger},c_{\kbs,\alpha,2}^{\dagger},\dots, c_{\kbs,\alpha,M}^{\dagger}]$, where $M$ is the number of `bands', which can refer to different orbitals, layers or sublattices. We take $\Upsilon$ to be Hermitian matrices and $v_{\qbs}$ is a positive semi-definite matrix for $\qbs\neq 0$, such that $v_{-\qbs} = v_{\qbs}^{\dagger}$. For $\qbs$ we only require that $v_{\qbs}$ is Hermitian. For instance, for a multilayer problem with Coulomb interactions, we may take 
$[v_{\qbs}]_{i,j} = \frac{2\pi}{\abs{\qbs}}e^{-d_{ij}\abs{\qbs}}(1-\delta_{\qbs,\zero})-{2\pi d_{ij}}\delta_{\qbs,\zero}$, where $d_{ij}$ is the separation between layers $i$ and $j$. For Moir\'e systems, $\kbs$ becomes quasi-momenta in the Moir\'e BZ, so that the index $j$ label higher BZs and the various $O$ correspond to momentum transfers $\qbs$, $\qbs + \Gbf_1$, $\qbs+\Gbf_2$,$\dots$ ;  where $\Gbf_1$ and $\Gbf_2$ are the Moi\'e reciprocal vectors. It is convenient to factorize the interaction as $v_{\qbs} = u_{\qbs}u_{\qbs}^\dagger$, where $u_{\qbs}$ is a $M$-by-$N_c$ matrix, where $N_c$ is the number of non-zero eigenvalues of $v_{\qbs}$, which can be interpreted as different `channels'. For example, in the multi-layer case we have $M$ different channels when we take the separation into account but when we set $d_{ij}$ equal to zero, there is only one channel. The latter situation is what we considered in the main text. 

At the non-interacting level, the susceptibility is diagonal in flavor and equal to 
\begin{gather}
    \Pi_{\alpha}(\qbs,\ii\omega)_{\mu\nu} = \chi_0(O_{\qbs,\mu,\alpha}^{\dagger},O_{\qbs,\nu,\alpha}|\ii\omega) \\
    = \frac{1}{\Area}\int_{\Omega}\sum_{\kbs}\Tr[ \Upsilon_{\mu}^{\dagger}\frac{1}{\ii\omega - h_{\kbs}}\Upsilon_{\nu}\frac{1}{\ii(\Omega+\omega) - h_{\kbs+\qbs}}]\\
    = \frac{1}{\Area}\int_{\Omega}\sum_{\kbs,\lambda,\lambda'}\Tr[ \Upsilon_{\mu}^{\dagger}\frac{P_{\kbs,\lambda}}{\ii\Omega - \ep_{\kbs,\lambda}}\Upsilon_{\nu}\frac{P_{\kbs+\qbs,\lambda'}}{\ii(\Omega+\omega) - \ep_{\kbs+\qbs,\lambda'}}]\\
    = \frac{1}{\Area}\sum_{\kbs,\lambda,\lambda'}
    \Tr[ \Upsilon_{\mu}^{\dagger}{P_{\kbs,\lambda}}\Upsilon_{\nu}{P_{\kbs+\qbs,\lambda'}}] \frac{n_{\kbs+\qbs,\lambda'}-n_{\kbs,\lambda}}{\ii \omega-(\ep_{\kbs+\qbs,\lambda'}-\ep_{\kbs,\lambda})}
\end{gather}
where $\chi_0(A,B|\ii\omega)$ is the non-interacting suceptiblity between operators $A$ and $B$ at imaginary frequency $\ii\omega$ and $h_{\kbs}=\sum_{\lambda}P_{\kbs,\lambda}\ep_{\kbs,\lambda}$ with $P_{\kbs,\lambda} = \ketbra{\kbs,\lambda}{\kbs,\lambda}$ are projectors. Similar to the main text, the `polarization bubble matrix' has the following properties 
\begin{enumerate}
    \item  $[\Pi_{\alpha}(\qbs,\ii\omega)]^\dagger=\Pi_{\alpha}(-\qbs,\ii\omega)=\Pi_{\alpha}(\qbs,-\ii\omega)$;
    \item If the state is invariant under $\gel$, then $\Pi_{\gel(\alpha)}(\qbs,\ii\omega)=\Pi_{\alpha}(\gel(\qbs),\ii\omega)$;
    \item $\Pi_{\alpha}+\Pi_{\alpha}^{\dagger}$ is a non-negative matrix.
\end{enumerate}
Properties 1 and 2 are derived as in the App.~\ref{app:BubbleProps:bare}. Property three, is equivalent to showing that for any complex vector $w_{\mu}$, $\Re[w_{\mu}^*\Pi_{\mu\nu}w_{\nu}]\geq 0$. To show the latter, note that $\Tr[WP_{\kbs,\lambda}W^{\dagger} P_{\kbs+\qbs,\lambda'}] = \abs{\mel{\kbs+\qbs,\lambda'}{W}{\kbs,\lambda}}^2\geq 0 $ for any matrix $W$. Then taking $W =\sum_{\mu} w_{\mu}\Upsilon^{\dagger}_{\mu}$, 
\begin{equation}
    \Re\{w_{\mu}^*\Pi_{\mu\nu}w_{\nu}\} = \frac{1}{\Area} \sum_{\kbs,\lambda,\lambda'}\Tr[W^{\dagger}P_{\kbs,\lambda}W^{\dagger} P_{\kbs+\qbs,\lambda'}] 
    \frac{-(\ep_{\kbs+\qbs,\lambda'}-\ep_{\kbs,\lambda})(n_{\kbs+\qbs,\lambda'}-n_{\kbs,\lambda})}{\omega^2+(\ep_{\kbs+\qbs,\lambda'}-\ep_{\kbs,\lambda})^2} \geq 0 .
\end{equation} 
For ease of notation, we split $\Pi = \Hsf +\ii \Asf$, where $\Hsf$ and $\Asf$ are Hermitian matrices. Property 3 states that $\Hsf$ is a positive semi-definite matrix.

We can evaluate the energy similarly to the one channel model by a coupling constant integration. Using the FH theorem again, leads to a expression similar to Eq.~\ref{eq:ExpvalV}:
\begin{gather}
    \expval{V} = \frac{1}{2\Area} \sum_{\qbs}\sum_{\mu,\nu} \left[\int_0^{\infty}\Re\{\sum_{\qbs,\mu,\nu} {v_{\qbs}^{\mu,\nu} }\chi(O_{\qbs,\mu}^{\dagger}, O_{\qbs,\nu}| \ii\omega)\}\frac{\dd\omega}{\pi} - \expval{v_{\qbs}^{\mu,\nu}\sum_{\alpha}\delta_{\qbs;\alpha}^{\mu,\nu}}\right]\\
    \delta_{\qbs;\alpha}^{\mu,\nu} =  
    \sum_{\kbs} c_{\kbs-\qbs,\alpha}^{\dagger}
    \Upsilon^{\mu} \Upsilon^{\nu} c_{\kbs-\qbs,\alpha}^{}.
\end{gather}

When $v_{\qbs}\neq 0$, the Hartree self-energy is generically non-trivial. The self-consistent equation for the self-energy is 
\begin{equation}
    [\Sigma^{\rm{H}}]_{i,j}(\kbs \alpha, \kbs'\alpha')  = \delta_{\kbs,\kbs'}\delta_{\alpha,\alpha'}\sum_{\beta,\mu,\nu} v_{\qbs=\zero}^{\mu,\nu} \frac{[(\Upsilon^{\mu})^\dagger]_{i,j}\expval{
    c^{\dagger}_{\kbs,\beta}
    \Upsilon^{\nu}c^{}_{\kbs,\beta}
    }+ (\mu\leftrightarrow \nu)}{2},
\end{equation}
where we assumed that the Hartree self-energy is diagonal in momentum and flavor. After solving the above equation, the Green's function of $\alpha$ fermions is  $\Gmc_{\alpha}^{\rm{H}}(\ii\omega,\kbs) = 1/(\ii\omega -h_{\kbs,\alpha} - \Sigma^{\rm{H}})$ and the polarization bubble should be evaluated with $h_{\kbs,\alpha} + \Sigma^{\rm{H}}$ instead of $h_{\kbs,\alpha}$. We denote this quantity by $\chi_0^{\rm{H}}$ instead of $\chi_0^{}$.

To make connection to the main text where we can forget about the Hartree self-energy, we write 
\begin{align}
    \expval{V}_{\rm{d}} &= \sum_{\mu,\nu}\frac{v_{\zero}^{\mu,\nu}}{2\Area} \left[\int_0^\infty \Re[\chi_0^{\rm{H}}(O_{\zero,\mu}^{\dagger},O_{\zero,\nu}^{}|\ii\omega)]\frac{\dd\omega}{\pi}- \expval{\delta_{\zero;\alpha}^{\mu,\nu}}\right],\\
    \expval{V}_{\rm{x}} &= \sum_{\mu,\nu}\sum_{\qbs\neq \zero}\frac{v_{\qbs}^{\mu,\nu}}{2\Area} \left[\int_0^\infty \Re[\chi_0^{\rm{H}}(O_{\qbs,\mu}^{\dagger},O_{\qbs,\nu}^{}|\ii\omega)]\frac{\dd\omega}{\pi}- \expval{\delta_{\qbs;\alpha}^{\mu,\nu}}\right],\\
    \expval{V}_{\rm{c}} &= \sum_{\mu,\nu}\sum_{\qbs}\frac{v_{\qbs}^{\mu,\nu}}{2\Area} \left[\int_0^\infty \Re[\chi(O_{\qbs,\mu}^{\dagger},O_{\qbs,\nu}^{}|\ii\omega)-\chi_0^{\rm{H}}(O_{\qbs,\mu}^{\dagger},O_{\qbs,\nu}^{}|\ii\omega)] \frac{\dd\omega}{\pi}\right],
\end{align}
where $\chi$ is the exact susceptibility. 
Note that $\expval{V}_{\rm{d}}$, $\expval{V}_{\rm{x}}$ and $\expval{V}_{\rm{c}}$ are the Hartree, Fock and correlation energies evaluated with respect to the Hartree self-consistent solution.

In TDH/RPA, the susceptibility is 
\begin{align}
    [\chi(\qbs,\ii\omega)]_{\mu\nu}  &\equiv \chi(O^{}_{\qbs,\mu}\,,\,O^{\dagger}_{\qbs,\nu}|\ii\omega) = \sum_{\kappa}[\chi_{0}(\qbs,\ii\omega)]_{\mu\kappa}\left(\frac{1}{1+v_{\qbs} \cdot \chi_{0}^{}(\qbs,\ii\omega)}\right)_{\kappa,\nu},\\
    [\chi_0^{}(\qbs,\ii\omega)]_{\mu\nu}&\equiv \chi_{0}^{\rm{H}}(O^{}_{\qbs,\mu}\,,\,O^{\dagger}_{\qbs,\nu}|\ii\omega).
\end{align}
As we are interested in the energy of the spin-polarized and valley-polarized half-metals, we simply restrict $\alpha$ to the two occupied values in the above expressions for each sector. Analogously to the main text, we assume that the self-consistent Hartree Green's functions for flavor $\alpha=1$ is equal to that of flavor $\alpha=2$ in the VP, and related by a symmetry to that of flavor $\alpha=3$ in the SP sector.  

In contrast to the main text, the coupling constant integration is no longer trivial because $\chi_0^{\rm{H}}$ depends implicitly on $\lambda$ via $v_{\qbs = \zero}$. However, we will show that $\expval{V}$ satisfies the same trends as in the main text.

Compared to the main text, there are two new complications: (1) the $\qbs =\zero$ components; and (2) the matrix structure of the correlation functions. The first complication is not really an issue because the correlations are evaluated at $\qbs=\zero$ which is a momentum invariant under any symmetry $g$. We can thus restrict to the $\qbs\neq \zero$ terms in following. 

Using the factorization $v_{\qbs} = u_{\qbs}u_{\qbs}^{\dagger}$, we can write
\begin{equation}
    \Tr[v_{\qbs}\chi(\qbs,\ii\omega)] = \Tr[ \frac{1}{1+ u_{\qbs}^{\dagger}\chi_{0}(\qbs,\ii\omega)u_{\qbs}}u_{\qbs}^{\dagger}\chi_{0}(\qbs,\ii\omega)u_{\qbs}] = 
    \Tr[1 - \frac{1}{1+ u_{\qbs}^{\dagger}\chi_{0}(\qbs,\ii\omega)u_{\qbs}}].
\end{equation}
Write $ u_{\qbs}^{\dagger}\chi_{0}(\qbs,\ii\omega)u_{\qbs} = \Xsf + \ii \Ysf$, for Hermitian matrices $\Xsf = u^{\dagger}\Hsf u \geq 0$ and $\Ysf=u^{\dagger}\Asf u$. 

For each case, we repeat the arguments around Eq.~\ref{eq:CaseI:Ineq} and Eq.~\ref{eq:CaseII:Ineq} in the main text to show the inequalities:

\def\Zsf{\mathsf{Z}}
\textit{Case I.--} In this case, $\Asf =0$. Furthermore, $[u^{\dagger}\chi_{0}u]_{\rm{SP}} = \Xsf_1 +  \Xsf_2$ and $[u^{\dagger}\chi_{0}u]_{\rm{VP}} = 2\Xsf_1$. To show that VP has lower energy than SP, we use that fact that the function $f(\Xsf) = \Tr[1 - (1+\Xsf)^{-1}]$ is a concave function in the space of positive semi-definite matrices $\Xsf$. 

To show this, we first calculate the partial derivatives in the open set $\mathcal{P}_{\epsilon}=\{\Xsf: \Xsf + \epsilon 1 >0 \,\,; \Xsf^\dagger=\Xsf\}$ with $0<\epsilon< 1$: $\grad_{a}\grad_{b}f = -\Tr[(1+\Xsf)^{-2}(\grad_a\Xsf)
(1+\Xsf)^{-1}(\grad_b\Xsf)]+ ( a\leftrightarrow b)$. The Hessian is non-negative if $-2\Tr[(1+\Xsf)^{-2} \Zsf (1+\Xsf)^{-1}\Zsf] \leq 0$ for any $\Zsf$ in the tangent space of Hermitian matrices. To show the latter, we expand in the eigenbasis of $\Xsf$: $\Xsf = \sum_{n} \ketbra{n}{n}x_n $ where $x_n >-\epsilon$ if $\Xsf\in \mathcal{P}_{\epsilon}$. Then
\begin{equation}
    -2\Tr[(1+\Xsf)^{-2} \Zsf (1+\Xsf)^{-1}\Zsf] = -2\sum_{n,n'} 
    (1+x_n)^{-2}
    (1+x_{n'})^{-1}
    \abs{\mel{n'}{\Zsf}{n}}^2\leq 0. 
\end{equation}
This shows that $f(\Xsf) = \Tr[1-(1+\Xsf)^{-1}]$ is a concave function in $\mathcal{P}_{\epsilon}$. Therefore, for any positive semi-definite matrices $\Xsf_1$ and $\Xsf_2$, 
\begin{equation}
    2f(\Xsf_1+\Xsf_2) \geq f(2\Xsf_1) + f(2\Xsf_2).
\end{equation}
From this follows that valley polarization is preferred over spin polarization. 

\textit{Case II.--} in this case $[u^{\dagger}\chi_{0}u]_{\rm{SP}} = \Xsf$ and $[u^{\dagger}\chi_{0}u]_{\rm{VP}} = \Xsf + \ii\Ysf$, with $\Xsf\geq 0 $ and $\Ysf^\dagger=\Ysf$. Note the following identity
\begin{equation}
    \Tr[\frac{1}{1+\Xsf + \ii\Ysf}] =  
    \Tr[\frac{1}{1+\Xsf} \frac{1}{1+\ii \Ssf}]; 
    \quad \Ssf= \frac{1}{\sqrt{1+\Xsf}} \Ysf \frac{1}{\sqrt{1+\Xsf}}.
\end{equation}
Using the fact that $\Ssf$ is Hermitian, we obtain the following identities
\begin{gather}
    \Re\Tr[v_{\qbs}\chi_{0}(\qbs,\ii\omega)]_{\rm{VP}}
    = \Tr[1-\frac{1}{1+\Xsf}\frac{1}{1+\Ssf^2}]\\
    \Re\Tr[v_{\qbs}\chi_{0}(\qbs,\ii\omega)]_{\rm{SP}}
    = \Tr[1-\frac{1}{1+\Xsf}]\\
    \Re\Tr[v_{\qbs}(\chi_{0}(\qbs,\ii\omega)_{\rm{SP}}-\chi_{0}(\qbs,\ii\omega)_{\rm{VP}})]
    = -\Tr[\frac{1}{\sqrt{1+\Xsf}}\frac{\Ssf^2}{1+\Ssf^2}\frac{1}{\sqrt{1+\Xsf}}]\leq0
\end{gather}

\end{document}